\documentclass[%
 aip,
 jmp,%
 amsmath,amssymb,
 reprint,%
]{revtex4-1}

\usepackage{xprintlen}

\usepackage{graphicx}
\usepackage[english]{babel}
\usepackage{microtype}          
\usepackage[breaklinks=true,colorlinks=true,linkcolor=blue,urlcolor=blue,citecolor=blue]{hyperref}

\usepackage{xcolor}     
\usepackage{mathtools}
\usepackage{braket}
\usepackage{tikz}
 \usepackage{printlen}

 \usepackage{longtable}
  \usepackage{booktabs}
 \usepackage{siunitx}
 \usepackage{csquotes}

\newcommand{\op}[1]{\ensuremath{\hat{#1}}}
\newcommand{\ladderdown}{\ensuremath{\op{a}^{\vphantom{\dagger}}}}
\newcommand{\ladderup}{\ensuremath{\op{a}^\dagger}}
\newcommand{\annihilop}{\ladderdown}
\newcommand{\creationop}{\ladderup}

\usepackage{dcolumn}
\usepackage{bm}

\begin{document}
\bibliographystyle{apsrev}
\preprint{APS/123-QED}

\title{Configuration Path Integral Monte Carlo Approach\\ to the Static Density Response of the Warm Dense Electron Gas}

\author{Simon Groth}
 \email{groth@theo-physik.uni-kiel.de}
\author{Tobias Dornheim}%
\author{Michael Bonitz}
\affiliation{Institut f\"ur Theoretische Physik und Astrophysik, Christian-Albrechts-Universit\"{a}t zu Kiel, D-24098 Kiel, Germany}

\date{\today}

\begin{abstract}
Precise knowledge of the static density response function (SDRF) of the uniform electron gas (UEG) serves as key input for numerous applications, most importantly for density functional theory beyond generalized gradient approximations. Here we extend the configuration path integral Monte Carlo (CPIMC) formalism that was previously applied to the spatially uniform electron gas to the case of an inhomogeneous electron gas by adding a spatially periodic external potential. This procedure has recently been successfully used in permutation blocking path integral Monte Carlo simulations (PB-PIMC) of the warm dense electron gas [Dornheim \textit{et al.}, Phys. Rev. E in press, arXiv:1706.00315], but this method is restricted to low and moderate densities. Implementing this procedure into CPIMC allows us to obtain exact finite temperature results for the SDRF of the electron gas at \textit{high to moderate densities} closing the gap left open by the PB-PIMC data. 
In this paper we demonstrate how the CPIMC formalism can be efficiently extended to the spatially inhomogeneous electron gas and present the first data points.
Finally, we discuss finite size errors involved in the quantum Monte Carlo results for the SDRF in detail and present a solution how to remove them that is based on a generalization of ground state techniques.

\end{abstract}

\maketitle

\section{Introduction}
The uniform electron gas (UEG) is one of the most important model systems of quantum physics and chemistry~\cite{loos,quantum_theory}. It is composed of electrons embedded in a uniform positive background -- to ensure charge neutrality. Thus, the UEG is well suited for thorough studies of physical effects induced by the long range Coulomb interaction of electrons in infinite quantum systems, such as collective excitations~\cite{pines,pines2} or the emergence of superconductivity~\cite{bcs}. The equilibrium state of the UEG is commonly determined by three parameters: (1)~the density (Brueckner) parameter $r_s=[3/(4\pi n)]^{1/3}/a_B$, with $a_B$ being the Bohr radius and $n$, the total density of spin-up and spin-down electrons, $n=n^\uparrow+n^\downarrow$; (2)~the degeneracy parameter $\theta=k_\text{B}T/E_\text{F}$, with the Fermi energy~\cite{fermi_E} $E_\text{F}$; and (3)~the spin-polarization, $\xi=(n^\uparrow -n^\downarrow)/n$, where, in this work, we focus on the most relevant case $\xi=0$, i.e., the unpolarized (paramagnetic) electron gas. Of particular current importance is the so-called ``warm dense matter'' regime~\cite{wdm_book} where the thermal energy is of the order of the Fermi energy ($\theta\sim 1$) while the densities are of the order of those found in solids ($r_s\sim 1$) or higher. Prominent examples for such extreme conditions are astrophysical applications~\cite{knudson,militzer}, dense quantum plasmas \cite{bonitz-book, bonitz_pre_13, michta_cpp_15},
inertial confinement fusion experiments~\cite{nora,schmit,hurricane3,kritcher}, or laser or ion beam excited solids~\cite{ernst, balzer_prb_16}.

The static density response function (SDRF), $\chi(\textbf{q})$, governs the density response to an external harmonic excitation of low amplitude $A$ and wave vector $\textbf{q}$, $\phi_{\mathbf{q}}(\mathbf{r})=2A\cos (\mathbf{r}\cdot\mathbf{q})$,
\begin{eqnarray}
\braket{\op{n}(\mathbf{r})}_A - \braket{\op{n}(\mathbf{r})}_0  =  \chi (\mathbf{q})\, \phi_{\mathbf{q}}(\mathbf{r})\;.
\nonumber
\end{eqnarray}
The SDRF (or longitudinal polarization function~\cite{bonitz-book}) is closely related to the static limit of the dielectric function and contains a wealth of information on the correlations and collective properties. 
Therefore, the SDRF is a key property of any correlated many-body system, for details, see Sec.~\ref{sec:lrt}. 

In particular, the SDRF
of the UEG at warm dense matter conditions constitutes a key ingredient for finite temperature density functional theory~\cite{wdm_book,mermin} (FTDFT) simulations within the adiabatic-connection fluctuation-dissipation formulation~\cite{lu,patrick,burke2}, the currently most promising way to improve DFT beyond the wide-spread generalized gradient approximation~\cite{dft_burke,pbe} and thereby enhance its predictive capabilities. In addition, the SDRF of the UEG can be used to directly compute the dynamic structure factor within the Born-Mermin-approach~\cite{fortmann1,fortmann2,collision1,collision2}, which is nowadays routinely measured for systems at warm dense matter conditions via X-ray Thomson scattering experiments. Moreover, knowledge of the exact SDRF of the UEG is highly useful for the computation of energy transfer rates~\cite{energy_transfer1,energy_transfer2}, electrical conductivity~\cite{cond1}, as well as for the construction of effective potentials~\cite{saum1,saum2,pseudo_potential, zhandos_pop_15}.

In the ground state, ab-initio results for the SDRF~\cite{moroni1,moroni2,bowen1,bowen2,senatore}, including a subsequent parametrization over a wide range of densities~\cite{cdop}, have been obtained long ago via diffusion Monte-Carlo simulations of the UEG subject to a weak periodic perturbation. However, even though the UEG effectively represents a one-component system, its simulation at warm dense matter conditions is highly challenging due to the fermion sign problem~\cite{loh,troyer} (FSP), which is particularly severe at finite temperature (cf.~Sec.~\ref{sec:fsp} for a detailed discussion of the FSP). Within the last years significant progress in this field could be achieved~\cite{dornheim_prl,groth2,dornheim_pop,groth_prl} via the introduction of two novel quantum Monte-Carlo (QMC) methods, which excel at complementary parameter regimes: permutation blocking path integral Monte-Carlo (PB-PIMC)~\cite{dornheim,dornheim2,dornheim3} is most efficient at low densities and strong coupling, whereas the configuration path integral Monte-Carlo (CPIMC) approach~\cite{tim1,tim_cpp15,tim3,groth} has no FSP at high densities, i.e., at weak coupling. Only recently, the PB-PIMC approach has been used to compute the first ab-initio results for the SDRF of the strongly coupled UEG at finite temperature~\cite{dornheim4}. However, these results are limited to density parameters of the order of $r_s=1$ and larger and cannot access the important regime of higher densities.

Therefore, in this work, we turn to the complementary CPIMC approach\cite{tim1} to compute the SDRF of the high density warm electron gas. This means, we extend the CPIMC formalism  from the homogeneous to the inhomogeneous electron gas such that it allows for the exact inclusion of an (in principle arbitrarily strong) periodic external potential. 
This allows us to obtain the first \textit{ab initio} data for the SDRF in the high-density regime ($r_s=0.5; 1$, $0.0625 \le \Theta \le 1$) and opens the way for systematic studies in the near future.

Moreover, since the simulations are restricted to finite systems with a few tens of electrons in a finite simulation volume $V$, we provide a detailed discussion of and a highly efficient solution to the problem of finite size errors involved in the SDRF. 
This is crucial because one is actually interested in the thermodynamic limit (TDL) properties,  $N\to\infty$ at $N/V=\text{const.}$. Finally, we compare our exact result for the SDRF in the TDL with dielectric approaches such as the random phase approximation~\cite{rpa_original} and the self-consistent scheme proposed by Singwi, Tosi, Land and Sj\"olander (STLS)~\cite{stls_original,stls}. 

This paper is structured as follows: in Sec.~\ref{sec:lrt} we briefly discuss the model Hamiltonian of the inhomogeneous electron gas and the basic linear response equations that are utilized for the computation of the SDRF. Thereafter, Sec.~\ref{sec:fsp} continues with a detailed introduction to the general quantum Monte-Carlo approach including the origin and consequences of the FSP, followed by the generalization of the CPIMC formalism to the inhomogenous electron gas in Secs.~\ref{sec:cpimc} and ~\ref{sec:estimator}. We proceed with a discussion of the CPIMC results for the SDRF of the ideal and non-ideal electron gas in Sec.~\ref{sec:ideal} and ~\ref{sec:interacting}. In Sec.~\ref{sec:finite_size}, finite size errors are investigated in detail, and an effective solution is presented to obtain the exact SDRF in the TDL from CPIMC simulations.

\section{Theoretical Basis of the CPIMC Approach to the Inhomogeneous Electron Gas}

\subsection{Linear Response Theory of the Uniform Electron Gas\label{sec:lrt}}
The model system of the unperturbed UEG consists of $N$ electrons in a finite volume $V=L^3$ subject to periodic boundary conditions, where a positive homogeneous background is assumed to ensure charge neutrality. The Hamiltonian of this system in Hartree atomic units reads
\begin{eqnarray} \label{eq:H}
\hat H_0 = -\frac{1}{2}\sum_{i=1}^{N}\nabla^2_i + \frac{1}{2}\sum_{i=1}^N\sum_{j\neq i}^N \Psi_\text{E}(\mathbf{r}_i, \mathbf{r}_j) + \frac{N}{2}\xi_\text{M} \;,
\end{eqnarray}
with $\Psi_\text{E}(\mathbf{r},\mathbf{s})$ being the Ewald pair potential and $\xi_\text{M}$ the Madelung constant, see, e.g., Ref.~\onlinecite{fraser}. For the purpose of computing the SDRF of the UEG we apply a weak periodic external potential of the form~\cite{moroni1,moroni2,bowen1,bowen2,senatore}
\begin{eqnarray}\label{eq:ext}
\op{H}_\text{ext}(A)&= \sum_{i=1}^N 2A\cos{(\op{\mathbf{r}}_i\cdot\mathbf{q})},
\end{eqnarray}
with $\mathbf{q}=\frac{2\pi}{L}\mathbf{m}$, $\mathbf{m}\in \mathbb{Z}^3$,
so that the (total) perturbed Hamiltonian is given by
\begin{eqnarray}\label{eq:tot_ham}
\op{H}_\text{A}&=\op{H}_0+\op{H}_\text{ext}(A)\;.
\end{eqnarray}
In the linear response regime, i.e., for sufficiently small amplitudes $A$, the induced density modulation is entirely determined by the SDRF~\cite{quantum_theory} $\chi$:
\begin{eqnarray}\label{eq:disturbed_density}
\braket{\op{n}(\mathbf{r})}_A - \braket{\op{n}(\mathbf{r})}_0  =  \chi (\mathbf{q})\, 2A\cos(\mathbf{r}\cdot\mathbf{q})\;,
\end{eqnarray}
where $\braket{\op{n}(\mathbf{r})}_0=n_0=\frac{N}{V}$ is the electron density of the unperturbed UEG. Hence, one may obtain $\chi(\mathbf{q})$ by computing the expectation value of the density operator $\op{n}(\mathbf{r})=\sum_{i=1}^N\delta(\mathbf{r}-\op{\mathbf{r}}_i)$
in the perturbed system and then fit the RHS of Eq.~(\ref{eq:disturbed_density}) to the LHS (see e.g.~Ref.~\onlinecite{dornheim4}). However, it turns out to be more convenient to compute $\chi$ directly from the Fourier transform of the density operator $\op{\rho}_\mathbf{q}=\frac{1}{V}\sum_{i=1}^N e^{-i\mathbf{q}\mathbf{r}_i}$ via the well-known relation~\cite{quantum_theory,bowen2}
\begin{eqnarray}\label{eq:chi}
\chi(\mathbf{q}) = \frac{1}{A}\braket{\op{\rho}_\mathbf{q}}_A\;.
\end{eqnarray}
In practice, we carry out several simulations for different amplitudes $A$ of the external field and then perform a linear fit to $\braket{\op{\rho}_\mathbf{q}}_A$ in dependence of $A$ where the resulting slope is $\chi$.

\subsection{Path integral Monte Carlo and the fermion Sign Problem\label{sec:fsp}}
Throughout this work we are interested in the computation of thermodynamic expectation values in the canonical ensemble, i.e., at fixed electron number $N$, volume $V$ and temperature $T$. For this task path integral Monte Carlo (PIMC) methods have proven to be a very powerful tool. The general idea of all existing PIMC approaches is to find a suitable expansion of the partition function of the form
\begin{eqnarray}\label{eq:Z}
Z=\text{Tr} e^{-\beta\op{H}} = \sum_{C} W(C)\;,
\end{eqnarray}
where $\beta=1/k_BT$ and $C$ denotes some high-dimensional multi-variable with an associated weight $W(C)\in \mathbb{R}$ that is readily evaluated. In the context of QMC, we commonly refer to $C$ as being a configuration. Given some concrete expansion of $Z$, thermodynamic expectation values of an arbitrary observable $\op{O}$ are written as
\begin{eqnarray}\label{eq:Expectation}
\braket{\op{O}} =\frac{1}{Z} \sum_{C} O(C)W(C)\;,
\end{eqnarray}
with $O(C)$ being the so-called estimator. If the weight function is strictly positive for a all configurations,  $W(C)>0$ $\forall$ $C$, such expressions can be efficiently computed via the Metropolis algorithm~\cite{metropolis}. The strength of this algorithm is that it allows to randomly sample configurations $\{C_0,C_1,\ldots,C_{N_C}\}$ with the correct probability $P(C)=\frac{1}{Z}W(C)$ without knowing the normalization constant $Z$. Starting from some initial configuration $C_0$ this is achieved by proposing a transition from $C_i$ to some randomly chosen $C^\prime$ and accepting this change, i.e., setting $C_{i+1}=C^\prime$, with the probability
\begin{eqnarray}\label{eq:acc_prob}
A(C\to C^\prime) = \text{min} \left\{1,\frac{W(C^\prime)}{W(C)}\right\}\;.
\end{eqnarray}
Having properly sampled the configurations in the described way, an asymptotically exact estimator of the expectation value Eq.~(\ref{eq:Expectation}) is immediately given by the average
\begin{eqnarray}
\braket{\op{O}}= \lim_{N_C\to\infty} \frac{1}{N_C}\sum_{i=1}^{N_C}O(C_i)\;.
\end{eqnarray}
In practice we are of course restricted to a finite number of sampled configurations $C_i$ so that the results are generally afflicted with a statistical uncertainty that can, in principle, be made arbitrarily small by increasing the  computation time (see Eq.~(\ref{eq:fsp})). Therefore, one may refer to Monte-Carlo methods as being ``quasi-exact''.

However, to this day there exists no exact expansion of the form Eq.~(\ref{eq:Z}) for fermionic quantum systems with a strictly positive weight function, and hence, it cannot be interpreted as a probability. To nevertheless utilize the Metropolis algorithm one can circumvent this issue by defining a modified (artificial) partition function
\begin{eqnarray}\label{eq:Z_prime}
Z^\prime= \sum_{C} |W(C)|\;,
\end{eqnarray}
and rewrite the expectation values as
\begin{eqnarray}\label{eq:rewritten_obs}
\langle O \rangle = \frac{\langle Os\rangle^\prime}{\langle s \rangle^\prime}\;, 
\end{eqnarray}
where $s=\text{sign}(W)$ so that 
\begin{eqnarray}
\braket{s}^\prime = \frac{1}{Z^\prime}\sum_{C} \text{sign}(W) |W(C)|=\frac{Z}{Z^\prime}\, ,
\end{eqnarray}
is simply the average sign of all sampled configurations in the modified configuration space. It is easy to see that the relative statistical uncertainty of expectation values computed in this way is inversely proportional to the average sign. Further, with $Z=e^{-\beta N f}$, where $f$ is the free energy per particle, it is
\begin{eqnarray}
\braket{s}^\prime = e^{-\beta N(f-f^\prime)}\;. 
\end{eqnarray}
Consequently, the relative statistical error of observables grows exponentially with the particle number $N$ and the inverse temperature $\beta$, while it can only be reduced with the square root of the number of generated samples $N_C$:
\begin{eqnarray}\label{eq:fsp}
\frac{\Delta O}{\braket{O}} \sim \frac{1}{\sqrt{N_C}}e^{\beta N(f-f^\prime)}\;.
\end{eqnarray}
This is the manifestation of the well-known fermion sign problem, which causes the simulation of fermions to be a highly demanding task even in thermodynamic equilibrium. Moreover, it has been shown that the sign problem is NP-hard~\cite{troyer}.

In the standard PIMC approach~\cite{cep}, the utilized expansion of the partition function is obtained by evaluating the trace in Eq.~(\ref{eq:Z}) in coordinate representation, leading to configurations $C$ that can be interpreted as paths or trajectories of all $N$ particles in imaginary time. In this formulation, the required anti-symmetrization of the density operator to correctly account for the Fermi statistics is the source of the sign changes in the weight function, and hence, of the FSP itself. Fortunately, the permutation blocking PIMC (PB-PIMC) method~\cite{dornheim,dornheim2,dornheim3}, developed by one of us, significantly reduces the FSP through a sophisticated rewriting of the partition function whereby paths with a different sign are combined into a single configuration. However, due to the formulation in coordinate representation, the PB-PIMC approach excels at strong coupling but still suffers from an increasing FSP towards lower temperature, preventing simulations of the UEG below half the Fermi temperature.

An alternative strategy, which is pursued in this paper, is given by the configuration path integral Monte Carlo (CPIMC) approach~\cite{tim_cpp15,tim1,tim3,groth}. In contrast to standard PIMC, this method is formulated in Fock-space, which leads to a FSP that is complementary to that of PB-PIMC: there is no sign problem at all for the ideal fermi gas but the FSP increases with coupling. For this reason, CPIMC has been highly valuable regarding the simulation of the (unperturbed) UEG at densities $r_s\lesssim1$, practically across the entire relevant temperature range~\cite{tim3}. In the next section, the CPIMC formalism will be generalized to the perturbed (inhomogeneous) electron gas described by the Hamiltonian Eq.~(\ref{eq:tot_ham}).

\subsection{CPIMC Approach to the Inhomogeneous Electron Gas\label{sec:cpimc}}
For the CPIMC formulation of the electron gas, we switch to second quantization with respect to plane wave spin orbitals $\langle \mathbf{r} \sigma \;|\mathbf{k}_i\sigma_i\rangle = \frac{1}{L^{3/2}} e^{i\mathbf{k}_i \cdot \mathbf{r}}\delta_{\sigma,\sigma_i}$ with $\mathbf{k}=\frac{2\pi}{L}\mathbf{m}$, $\mathbf{m}\in \mathbb{Z}^3$ and $\sigma_i\in\{\uparrow,\downarrow\}$.  
The $N$-particle states are then given by Slater determinants in Fock space
\begin{eqnarray}\label{eq:fock_state}
|\{n\}\rangle=|n_1, n_2, \dots\rangle\;,
\end{eqnarray}
with the fermionic occupation number $n_i\in\{0,1\}$ of the $i$-th plane wave spin-orbital naturally satisfying $\sum_i n_i=N$. 
In this representation, the Hamiltonian is expressed in terms of the creation ($\op{a}^\dagger_i$) and annihilation ($\op{a}_i$) operators, which, when acting on the states [Eq.~(\ref{eq:fock_state})], create or annihilate a particle in the spin-orbital $i$. These operators satisfy the usual fermionic anti-commutation relations, thereby automatically incorporating the correct Fermi statistics.
The UEG Hamiltonian Eq.~(\ref{eq:H}) takes the explicit form~\cite{quantum_theory} 
\begin{eqnarray}\label{eq:H_second}
\op{H}_0=
\frac{1}{2}\sum_{i}\mathbf{k}_i^2 \creationop_{i}\annihilop_{i} + \smashoperator{\sum_{\substack{i<j,k<l \\ i\neq k,j\neq l}}} 
w^-_{ijkl}\creationop_{i}\creationop_{j} \annihilop_{l} \annihilop_{k} + N\frac{\xi_\text{M}}{2}\;,\quad
\end{eqnarray}
with the antisymmetrized two-electron integrals $w^-_{ijkl} =w_{ijkl}-w_{ijlk}$, where
\begin{eqnarray}\label{eq:two_ints}
w_{ijkl}=\frac{4\pi e^2}{L^3 (\mathbf{k}_{i} - \mathbf{k}_{k})^2}\delta_{\mathbf{k}_i+\mathbf{k}_j, \mathbf{k}_k + \mathbf{k}_l}\delta_{\sigma_i,\sigma_k}\delta_{\sigma_j,\sigma_l}\;.
\end{eqnarray}
Likewise, for the external potential Eq.~(\ref{eq:ext}) we have
\begin{eqnarray}\label{eq:H_ext_second}
\op{H}_\text{ext} = \sum_{i\neq j}a_{ij}\, \op{a}^\dagger_i\op{a}_i\;,
\end{eqnarray}
with the corresponding one-electron integrals
\begin{eqnarray}\label{eq:1p_matrix_element}
a_{ij} = A \delta_{\sigma_i\sigma_j}(\delta_{\mathbf{k}_j-\mathbf{k}_i,\mathbf{q}} + \delta_{\mathbf{k}_j-\mathbf{k}_i,-\mathbf{q}})\;.
\end{eqnarray}
The main idea of CPIMC is to split the total Hamiltonian into an off-diagonal ($\op{Y}$) and diagonal part ($\op{D}$) with rsepect to the Fock states, Eq.~(\ref{eq:fock_state}), so that $\op{H}_A=\op{H}_0+\op{H}_\text{ext}=\op{D}+\op{Y}$.
The matrix elements of these operators are readily computed according to the well-known Slater-Condon rules \cite{tim1}
\begin{align}\label{eq:matrix_elements}
 D_{\{n\}} &= \frac{1}{2}\sum_l \mathbf{k}_l^2 n_{l} + \frac{1}{2}\sum_{l<k}w^-_{lklk}n_{l}n_{k}\;,\\
Y_{\{n\},\{\bar{n}\} }&=\begin{cases}
   a_{ij}(-1)^{\alpha_{\{n\},pq}}, &\{n\}=\{\bar{n}\}_q^p\\[0.2cm]
w_{pqrs}^- (-1)^{\alpha_{\{n\},pq}+\alpha_{\{\bar{n}\},rs}}, & \{n\}=\{\bar{n}\}_{r<s}^{p<q}\nonumber
\end{cases}
\end{align}  
with the fermionic phase factor
\begin{eqnarray}\label{eq:phase_factor}
\alpha_{\{n\},pq} &= \displaystyle\sum_{l=\min(p,q)+1}^{\max(p,q)-1}n_l\;.
\end{eqnarray}
The notation $\{n\}_q^p$ describes an excitation from an occupied orbital $q$ to a free orbital $p$ in the state $\ket{\{n\}}$. Hence, we observe that there are only two possibilities for non-vanishing off-diagonal elements: the states $\ket{\{n\}}$ and $\ket{\{\bar{n}\}}$ can differ in either exactly two $(pq)$ or four orbitals $(pqrs)$. This is a direct consequence of the fact that the Hamiltonian only contains strings of two or four creation and annihilation operators. For completeness, we mention that for the general case of an arbitrary system Hamiltonian there is an additional contribution to the off-diagonal elements where $\{n\}=\{\bar{n}\}_q^p$:
\begin{eqnarray}
Y_{\{n\},\{\bar{n}\}} = \sum_{\substack{i=0\\i\neq p,q}}w^-_{ipiq} n_i (-1)^{\alpha_{\{n\},pq}}\;.
\end{eqnarray}
For the electron gas this contribution vanishes since here the two-particle integrals with only two equal indices are always zero due to the Kronecker delta in Eq.~(\ref{eq:two_ints}), which ensures that the total momentum of the two particles before and after the excitation is conserved.

After having split the Hamiltonian into its diagonal and off-diagonal part, we switch to the interaction picture in imaginary time with respect to $\op{D}$ and make use of the identity:
\begin{align}
 e^{-\beta\op{H}}&=e^{-\beta\op{D}}\op{T}_\tau e^{-\int_0^\beta\op{Y}(\tau)\mathrm{d}\tau}\nonumber\;,\\ 
 \op{Y}(\tau)&=e^{\tau\op{D}}\op{Y}e^{-\tau\op{D}}\;,
\end{align}
with $\op{T}_\tau$ being the time-ordering operator. Plugging this identity into Eq.~(\ref{eq:Z}) and computing the trace using the Slater determinants, Eq.~(\ref{eq:fock_state}), finally yields \cite{tim1}
\begin{align} \label{eq:Z_expansion}
Z = &
\sum_{K=0,\atop K \neq 1}^{\infty} \sum_{\{n\}}
\sum_{s_1\ldots s_{K-1}}\,
\int\limits_{0}^{\beta} d\tau_1 \int\limits_{\tau_1}^{\beta} d\tau_2 \ldots \int\limits_{\tau_{K-1}}^\beta d\tau_K 
 \\\nonumber
& (-1)^K  
e^{-\sum\limits_{i=0}^{K} D_{\{n^{(i)}\}} \left(\tau_{i+1}-\tau_i\right) } 
\prod_{i=1}^{K} Y_{\{n^{(i)}\},\{n^{(i-1)}\} }(s_i)\;.
\end{align}
Here, we have introduced the multi-index $s_i$ which defines the two or four orbitals in which the states $\ket{\{n^{(i)}\}}$ and $\ket{\{n^{(i-1)}\}}$ differ, i.e. $s_i=(pq)$ or $s_i=(pqrs)$. Further, all non-vanishing contributions in Eq.~(\ref{eq:Z_expansion}) obey the condition $\{n\}=\{n^{(0)}\}= \{n^{(K)}\}$. This way we have transformed the partition function, Eq.~(\ref{eq:Z}), into an exact infinite perturbation expansion with respect to the off-diagonal part of the Hamiltonian.

Comparing Eq.~(\ref{eq:Z_expansion}) with Eq.~(\ref{eq:Z}) we straightforwardly identify the multi-variable $C$ of each configuration contributing to the partition function:
\begin{eqnarray}
C=(K,\{n\}, s_1,\ldots,s_{K-1},\tau_1,\ldots,\tau_K)
\end{eqnarray}
with the corresponding weight function
\begin{align}\label{eq:weight}
W(C) =
(-1)^K \, & e^{-\sum\limits_{i=0}^{K} D_{\{n^{(i)}\}} \left(\tau_{i+1}-\tau_i\right) } \\
&\times \prod_{i=1}^{K} Y_{\{n^{(i)}\},\{n^{(i-1)}\} }(s_i)\;.\nonumber
\end{align}
\begin{figure}
\includegraphics[width=0.45\textwidth]{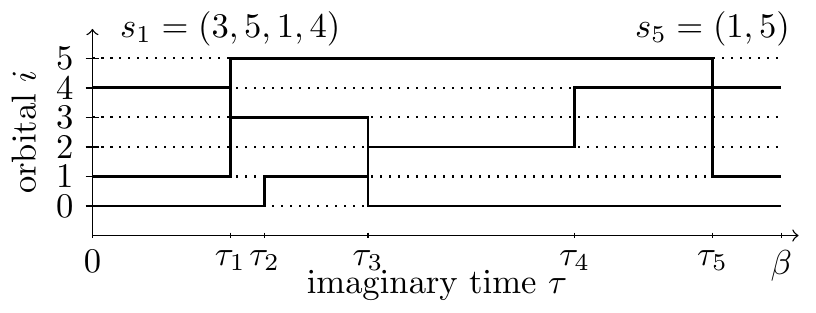}
\caption{\label{fig:example_path}Typical ``path'' in a CPIMC simulation of $N=3$ particles: the starting Slater determinant at time $\tau_0=0$ with the set of occupation numbers $\{n\}=\{110010\ldots\}$ undergoes five different one- or two-particle excitations of type $s_i$ at times $\tau_i$, $i=1\dots 5$.} 
\end{figure}
\begin{figure}
\includegraphics[width=0.3\textwidth]{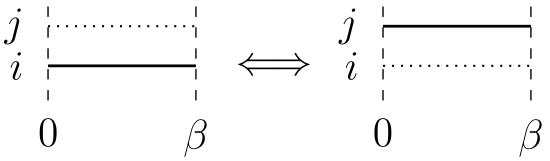}
\caption{\label{fig:diagram_excite_orbital}Diagram for exciting a whole occupied orbital $i$ (from $\tau=0$ to $\tau=\beta$) to an unoccupied orbital $j$.}
\end{figure}
\begin{figure}
\includegraphics[width=0.3\textwidth]{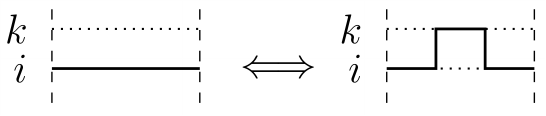}
\includegraphics[width=0.3\textwidth]{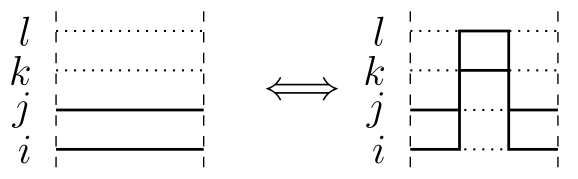}
\caption{\label{fig:diagram_add_pair_T2T4}Diagrams for adding a pair of type 2 (top) or type 4 kinks (bottom) via a one- or two-particle excitation, respectively.}
\end{figure}
Each configuration $C$ can be visualized as a $\beta-$periodic ``path in imaginary time''. But in contrast to standard PIMC which is formulated in coordinate space, here the path proceeds in Fock space and can be understood as follows: starting from an initial set of occupation numbers $\{n\}$ at $\tau_0=0$ one subsequently applies one- or two-particle excitations at times $\tau_i$, where the involved orbitals are defined by the multi-index $s_i$. An example of a typical path for a system of $N=3$ particles is shown in Fig.~\ref{fig:example_path}. 

According to the number of involved orbitals we refer to one- and two-particle excitations as ``kinks'' of type 2 and 4, respectively. Hence, in CPIMC, one randomly samples all possible closed paths with their associated weight, i.e., the modulus of Eq.~(\ref{eq:weight}), and computes observables via Eq.~(\ref{eq:rewritten_obs}). This is achieved by a highly complex set of Monte Carlo steps in which one proposes to add, remove, and change a single kink or pairs of kinks and accept or reject those changes with the Metropolis acceptance probability Eq.~(\ref{eq:acc_prob}). Starting from an initial path without kinks one can propose three changes: 1) one can simply excite a whole occupied orbital (from $\tau=0$ to $\tau=\beta$), which is illustrated in in Fig.~\ref{fig:diagram_excite_orbital}. 2) one can propose to add a pair of type 2 kinks or 3) a pair of type 4 kinks via a one- or two-particle excitation (see Fig.~\ref{fig:diagram_add_pair_T2T4}). Adding a single kink is not possible since this would violate the condition of $\beta-$periodicity, $\{n^{(0)}\}=\{n^{(K)}\}$. 

Once one has successfully added a pair of kinks, one can also add a single kink by changing another. A careful analysis reveals that there are in total 14 elementary diagrams for adding a single kink via a one- or two-particle excitation, which are all depicted in Fig.~\ref{fig:diagram_add_T2T4} in the Appendix. Naturally, to maximize the efficiency of the CPIMC simulation one only proposes to add such kinks that are associated with a non-vanishing off-diagonal matrix element, Eq.~(\ref{eq:matrix_elements}), i.e., which have a non-vanishing one- or two-electron integral. For example, when randomly choosing the two orbitals $q$ and $p$ for a one-particle excitation, one ensures that $|\mathbf{k}_p-\mathbf{k}_q|=|\mathbf{q}|$ with $\mathbf{q}$ being the wave vector of the periodic external potential, Eq.~(\ref{eq:H_ext_second}). Likewise, whenever proposing to add a type 4 kink one makes sure that momentum conservation is fulfilled.  

Finally, we point out that the major difference between the previous CPIMC formulation for the (unperturbed) UEG~\cite{tim_cpp15} and the present extension to the inhomogeneous electron gas lies in the occurrence of type 2 kinks (one-particle excitations), which are solely induced by the one-particle matrix elements $a_{ij}$
of the external potential in Eq.~(\ref{eq:matrix_elements}). In case of the UEG, $a_{ij}=0$, and hence, there are only momentum conserving type 4 kinks. This causes a large simplification of the algorithm since the 14 elementary diagrams of adding a single kink (see Fig.~\ref{fig:diagram_add_T2T4}) reduce to only three, i.e., those containing solely type 4 kinks. 

\subsection{CPIMC Estimator for the Static Responce Function\label{sec:estimator}}
\begin{figure*}
\includegraphics[width=0.4\textwidth]{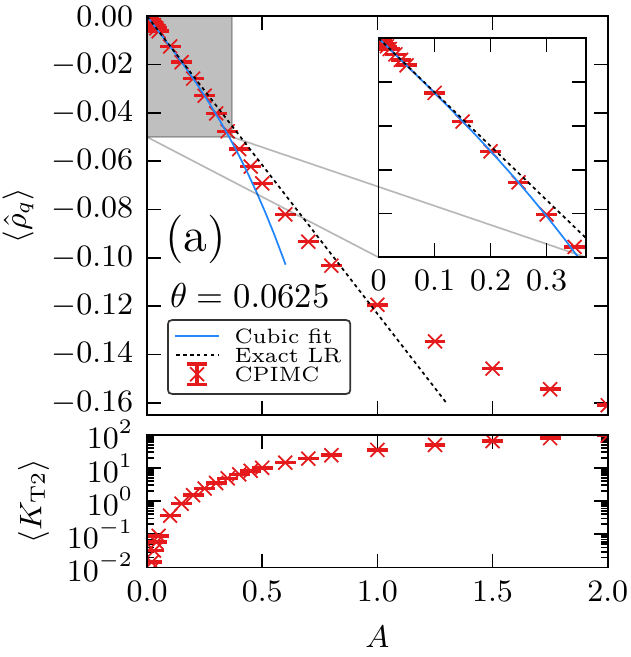}
\hspace*{0.5cm}
\includegraphics[width=0.4\textwidth]{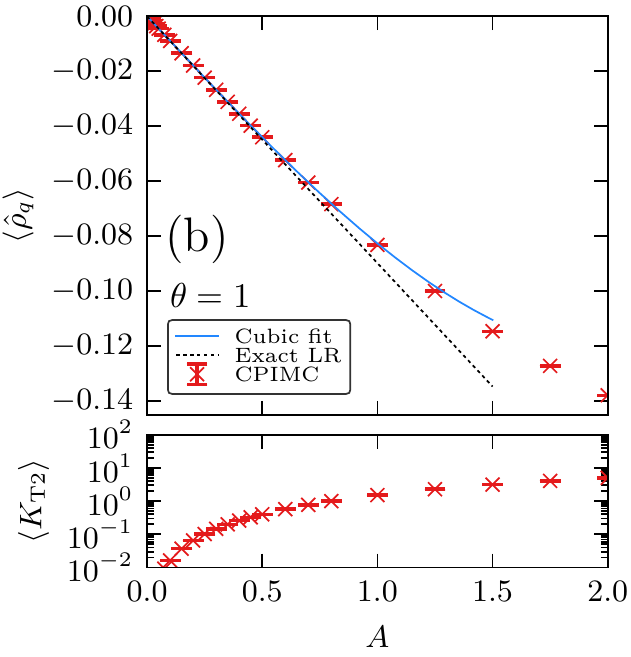}
\caption{\label{fig:ideal_Aext} \textbf{Top panels:} Dependence of the induced density $\braket{\op{\rho}_q}$ for $\mathbf{q}=\frac{2\pi}{L}(1,0,0)^{\text{T}}$ on the amplitude of the external field. Shown are CPIMC results (red crosses) for the ideal electron gas with $N=4$ electrons at $r_s=1$ for two different temperatures: (a) $\theta=0.0625$ and (b) $\theta=1$ (right). The blue curve represents a fit of Eq.~(\ref{eq:cubic_fit}) to the CPIMC data. The black dotted line corresponds to the exact LR behavior computed from Eq.~(\ref{eq:ideal_chi_formula}). \textbf{Bottom panels:} Dependence of the average number of type 2 kinks on the amplitude of the external field.
}
\end{figure*}
To compute the SDRF with CPIMC via Eq.~(\ref{eq:chi}), we need to derive an estimator for the Fourier transform of the density operator, $\op{\rho}_\mathbf{q}$, in correspondence to the CPIMC expansion of the partition function Eq.~(\ref{eq:Z_expansion}), i.e., we have to write its expectation value in the form of Eq.~(\ref{eq:Expectation}). Taking into account that $\braket{\op{\rho}_\mathbf{-q}} = \braket{\op{\rho}_\mathbf{q}}$, its second quantization representation is given by
\begin{eqnarray}\label{eq:rho_expec}
\braket{\op{\rho}_\mathbf{q}}=
\frac{1}{2V}\sum_{i\neq j} \delta_{\sigma_i\sigma_j}(\delta_{\mathbf{k}_j-\mathbf{k}_i,\mathbf{q}} + \delta_{\mathbf{k}_j-\mathbf{k}_i,-\mathbf{q}})\braket{\op{a}^\dagger_i\op{a}_j}\;,\;\;\;
\end{eqnarray}
and we immediately see that it can be computed directly from the off-diagonal elements of the one-particle density matrix $\braket{\op{a}^\dagger_p\op{a}_q}$. An estimator for these elements is readily obtained by using the relation
\begin{eqnarray}
\braket{\op{a}^\dagger_p\op{a}_q} = \frac{1}{Z}\text{Tr}\left\{\op{a}^\dagger_p\op{a}_q e^{-\beta\op{H}}\right\} 
= -\frac{1}{\beta}\frac{1}{Z} \frac{\partial Z}{\partial a_{pq}}\;,
\end{eqnarray}
and carrying out the derivative with the CPIMC expansion of the partition function, Eq.~(\ref{eq:Z_expansion}). This yields
\begin{align} \label{eq:1p_estimator}
\braket{\op{a}^\dagger_p\op{a}_q} = 
  \frac{1}{Z}\sum_{C}\left(-\frac{1}{\beta}\sum_{i=1}^{K}\frac{1}{a_{pq}}\delta_{s_i,(pq)}\right)W(C)\;,
\end{align}
where the abbreviation 
\begin{eqnarray}
\sum_{C}:= \sum_{K=0,\atop K \neq 1}^{\infty} \sum_{\{n\}}
\sum_{s_1\ldots s_{K-1}}\,
\int\limits_{0}^{\beta} d\tau_1 \int\limits_{\tau_1}^{\beta} d\tau_2 \ldots \int\limits_{\tau_{K-1}}^\beta d\tau_K \quad
\end{eqnarray}
has been used. By inserting Eqs.~(\ref{eq:1p_estimator}) and (\ref{eq:1p_matrix_element}) into (\ref{eq:rho_expec}) the estimator reduces to
\begin{eqnarray}\label{eq:estimator}
\braket{\op{\rho}_\mathbf{q}}= 
  \frac{1}{Z}\sum_{C}\left(-\frac{1}{2V\beta}\sum_{i=1}^{K}\frac{1}{A}\delta_{s_i,s_{\text{\tiny T2}}}\right)W(C)\;,
\end{eqnarray}
where $\delta_{s_i,s_\text{\tiny T2}}$ ensures that only those kinks contribute which are of type 2. Simply speaking, we just have to average over the number of type 2 kinks in all sampled paths and divide by $-2V\beta A$.

\section{CPIMC Simulation Results\label{sec:results}}

\subsection{Ideal Electron Gas\label{sec:ideal}}

Besides being highly valuable for the finite size correction of the SDRF discussed in Sec.~\ref{sec:finite_size}, the ideal Fermi system constitutes the natural first test case for CPIMC due to its formulation as an exact perturbation expansion in second quantization. It is realized by setting all two-particle matrix elements Eq.~(\ref{eq:two_ints}) to zero. In case of the (unperturbed) UEG there are, consequently, no kinks at all, so that the weight function [Eq.~(\ref{eq:weight})] is always positive, meaning that the average sign is always one. However, in simulations of the perturbed ideal electron gas, the sampled paths contain type 2 kinks induced by the external field, where each of them may cause up to two sign changes in the weight function Eq.~(\ref{eq:weight}) through: (1) the factor $(-1)^{K}$ and (2) the phase factor Eq.~(\ref{eq:phase_factor}) occurring in its matrix element Eq.~(\ref{eq:matrix_elements}). Yet, the average sign still remains unity. This is because, in the absence of type 4 kinks, type 2 kinks can only be added and removed in symmetric pairs as shown in Fig.~\ref{fig:diagram_add_pair_T2T4}---this is a simple consequence of the fact that all type 2 kinks $s=(pq)$ must fulfill $|\mathbf{k}_p-\mathbf{k}_q|=|\mathbf{q}|$. The induced sign changes of such pairs exactly compensate each other so that the strict positive definiteness of the weight function remains preserved, and hence, the FSP remains absent, in striking contrast to standard PIMC in coordinate space.   

As a first demonstration, we perform CPIMC simulations of the unpolarized ideal electron gas at $r_s=1$ with $N=4$ particles for different amplitudes $A$ of the external field with a wave vector $\mathbf{q}=\frac{2\pi}{L}(1,0,0)^{\text{T}}$. Fig.~\ref{fig:ideal_Aext} shows the results for the induced density $\braket{\op{\rho}_\mathbf{q}}$ (top) and the average number of type 2 kinks (bottom) in dependence of the amplitude for two different temperatures $\theta=0.0625$ (left) and $\theta=1$ (right). As a cross-check, the dotted black line has been computed from the unperturbed ideal UEG according to Eq.~(\ref{eq:ideal_chi_formula}) as discussed in Sec.~\ref{sec:finite_size}. In the linear response regime, both results must coincide, which is observed for $A\lesssim 0.2$ at $\theta=0.0625$, while at $\theta=1$ the linear response regime remains valid for much larger amplitudes, i.e., up to $A\sim 0.5$. This behaviour is also reflected in the average number of type 2 kinks for the same amplitude which is reduced by about two orders of magnitude at $\theta=1$ compared to $\theta=0.0625$. Interestingly, in both cases the linear regime is reached where $\braket{K_\text{T2}}\lesssim 1$. In addition, since the next order beyond the linear regime is given by the cubic response function~\cite{moroni2} $\chi^{(3)}$, we also perform a cubic fit (blue line) of the form
\begin{eqnarray}\label{eq:cubic_fit}
\braket{\op{\rho}_q}=\chi(q)A + \chi^{(3)}(q)A^3
\end{eqnarray}
to the CPIMC data up to $A=0.25$, for $\theta=0.0625$ and $A=0.5$, for $\theta=1$, respectively. Clearly, also the cubic regime remains valid for much larger amplitudes at higher temperatures.
\subsection{Interacting Electron Gas\label{sec:interacting}}
\begin{figure*}
\includegraphics[width=0.4\textwidth]{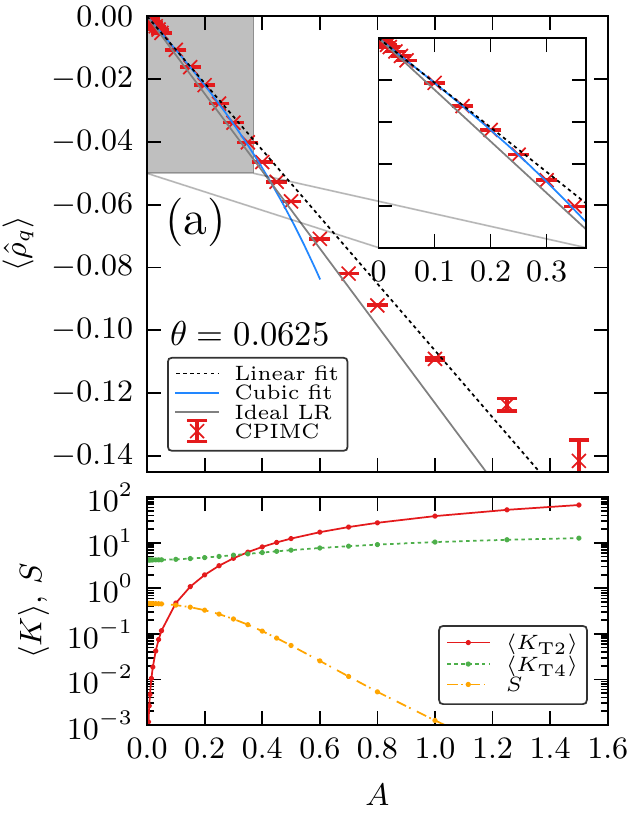}
\hspace*{0.5cm}
\includegraphics[width=0.4\textwidth]{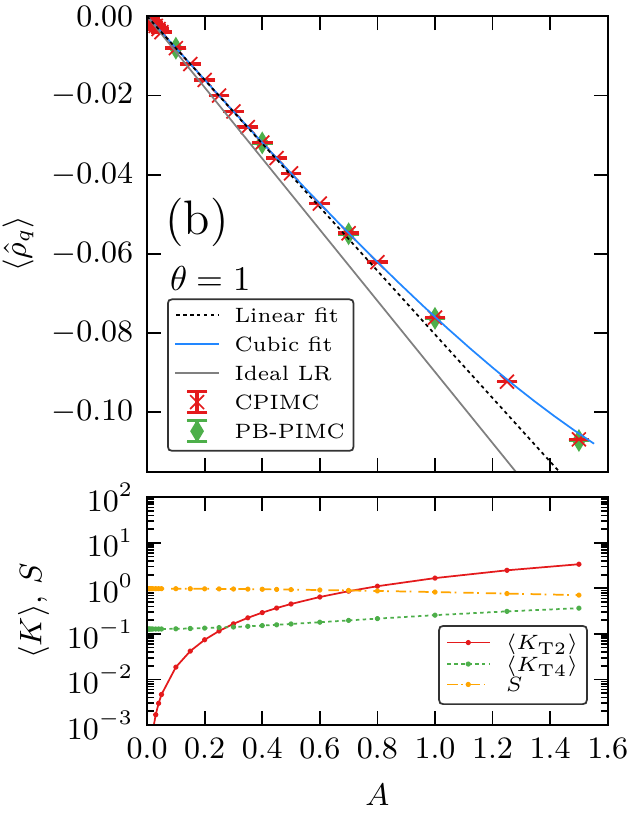}
\caption{\label{fig:interacting_Aext} 
\textbf{Top panels:} Dependence of the induced density $\braket{\op{\rho}_q}$ for $\mathbf{q}=\frac{2\pi}{L}(1,0,0)^{\text{T}}$ on the amplitude of the external field. Shown are CPIMC results (red crosses) for the interacting electron gas with $N=4$ electrons at $r_s=1$ for two different temperatures: (a) $\theta=0.0625$ and (b) $\theta=1$. The blue (black dotted) curve represents a cubic (linear) fit (cf.~Eq.~(\ref{eq:cubic_fit})) to the CPIMC data. The grey solid line shows the ideal LR behavior computed from Eq.~(\ref{eq:ideal_chi_formula}). For comparison, at $\theta=1$, we also plot  the PB-PIMC results (green diamonds). \textbf{Bottom panels:} Dependence of the average number of type 2 kinks (red), type 4 kinks (green), and the average sign (orange) on the amplitude of the external field.
}
\end{figure*}
\begin{figure}[h]
\includegraphics[width=0.4\textwidth]{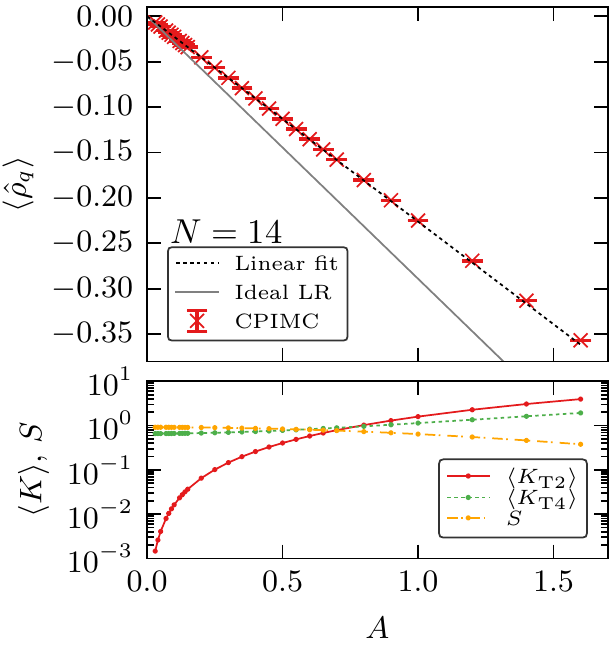}
\caption{\label{fig:interacting_Aext_N14}\textbf{Top panel:} Dependence of the induced density $\braket{\op{\rho}_\mathbf{q}}$ for $\mathbf{q}=\frac{2\pi}{L}(1,0,0)^{\text{T}}$ on the amplitude of the external field. Shown are CPIMC results (red crosses) for $N=14$ electrons at $r_s=0.5$ and $\theta=0.5$. The dotted black line corresponds to a linear fit. For comparison, we also plot the ideal LR behaviour (grey solid line). \textbf{Bottom panel:} Dependence of the average number of type 2 kinks (red), type 4 kinks (green), and the average sign (orange) on the amplitude of the external field.}
\end{figure}
Next, we perform the same CPIMC simulations for the interacting system (identical system parameters) as for the ideal case discussed in Section~\ref{sec:ideal}. The results are shown in Fig.~\ref{fig:interacting_Aext} where a linear (dotted black) and a cubic fit (solid blue) to the CPIMC data are depicted. For these parameters, we observe that the range of amplitudes for which the linear and cubic response regimes are valid is similar to that found for the ideal system. This is because the response of $N=4$ particles at $r_s=1$ is comparable to that of the ideal system (grey line). In addition, in the bottom panels the average number of type 4 kinks (green curve) is depicted, which are induced solely by the Coulomb interaction and which cause the average sign (orange curve) to deviate from one. In the linear regime, the dependence of the number of type 4 kinks on the amplitude is negligible. 

However, for larger values of $A$ not only the average number of type 2 kinks becomes very large but also the number of type 4 kinks increases significantly. The main reason for this behavior is the substantial increase of the configuration space with increasing amplitude. In particular, at $\theta=0.0625$ (left graphic) the average sign drops below $10^{-3}$ at $A>1$ and, according to Eq.~(\ref{eq:fsp}), the statistical error of the corresponding CPIMC results is clearly enhanced. As a further cross-check of the correctness of the presented algorithm, at $\theta=1$ we also compare with the PB-PIMC method (green diamonds), which are in perfect agreement with CPIMC, as expected.

In Fig.~\ref{fig:interacting_Aext_N14}, a similar investigation is carried out for a larger system containing $N=14$ electrons at $r_s=0.5$ and $\theta=0.5$. For these system parameters, the average sign (orange curve in the bottom panel) does not drop below $0.1$, even up to values of the amplitude $A \sim 1.5$. Thus, very precise CPIMC results for the induced density can be obtained. In comparison to the smaller system of $N=4$ electrons in Fig.~\ref{fig:interacting_Aext} the linear response regime is valid up to about twice as large amplitudes so that the SDRF $\chi$, given by the slope of the linear fit (dotted black line), can be obtained with a relative accuracy of up to $0.02\%$. Further, we observe that the average number of type 2 kinks $\braket{K_\text{T2}}$ (red curve in the bottom panel) is significantly larger than one for amplitudes $A>1.5$, and still, the deviation from the LR behaviour is only minor. Recalling that, for the smaller $N=4$ system in Fig.~\ref{fig:interacting_Aext}, the LR regime is valid for $\braket{K_\text{T2}}\lesssim 1$ we conclude that the average number of type 2 kinks alone is not a reliable indicator for the validity of the linear response regime.

\begin{figure}
\includegraphics[width=0.4\textwidth]{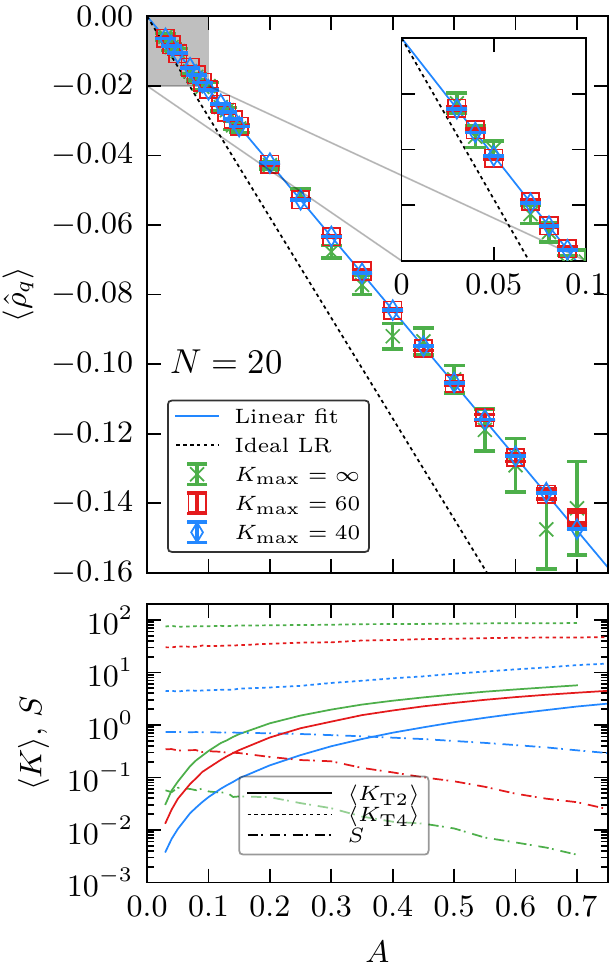}
\caption{\label{fig:interacting_Aext_N20}\textbf{Top panel:} Dependence of the induced density $\braket{\op{\rho}_\mathbf{q}}$ for $\mathbf{q}=\frac{2\pi}{L}(1,0,0)^{\text{T}}$ on thfe amplitude of the external field. Shown are CPIMC results for $N=20$ electrons at $r_s=0.5$ and $\theta=0.5$, where the maximum total number of kinks in the sampled paths has been restricted to $K_\text{max}=40$ (blue), $K_\text{max}=60$ (red), and $K_\text{max}=\infty$, i.e.~no restriction (green). The solid blue line corresponds to a linear fit to the data for $K_\text{max}=40$. The black dotted line shows the ideal LR behaviour. \textbf{Bottom panel:} Dependence of the average number of type 2 kinks (solid lines), type 4 kinks (dotted lines), and the average sign (dash-dot lines) on the amplitude of the external field. The colors correspond to the restrictions on the maximum number of kinks as labeled in the top panel.}
\end{figure}

When further increasing the system size to $N=20$, while keeping the density and degeneracy parameters unchanged at $r_s=0.5$ and $\theta=0.5$, the CPIMC simulations become significantly more demanding. This is demonstrated in Fig.~\ref{fig:interacting_Aext_N20}, where we artificially restricted the simulation to those configurations containing a maximum of 40 (blue), 60 (red), or arbitrarily many (green) kinks. More precisely, once a path with $K=K_\text{max}$ kinks is realized, we do not propose to add any further kinks. First, for the result obtained without any restrictions (green), we see that these data are afflicted with a clearly visible statistical noise, which is due to an average sign (bottom panel, dash-dot) that is smaller than $0.1$ even in the homogeneous case ($A=0$). Naturally, the resulting value for the SDRF from a linear fit to these data (not depicted) would only be of very poor quality. However, by restricting the total maximum number of kinks (blue and red curves), the average number of kinks (bottom panel, solid and dotted lines) is reduced by an order of magnitude, whereby the average sign is increased by an order of magnitude (dash-dotted lines). 

Normally, one would expect this procedure to bias the result for the density response since by imposing these restrictions, one only samples paths from a small region of the total configuration space. Instead, one observes that, within statistical error bars, all three simulations are in perfect agreement, both for large and small amplitudes $A$ (see inset in the upper panel). This very favourable behaviour is explained by a complete cancellation of all contributions from paths with a number of kinks  larger than the maximum. In other words, due to the sign changes in the weight function Eq.~(\ref{eq:weight}) the expansion of the physical partition function Eq.~(\ref{eq:Z_expansion}) converges for much smaller values of $K$ than the simulated primed partition function Eq.~(\ref{eq:Z_prime}). 
\begin{figure*}
\includegraphics[width=0.4\textwidth]{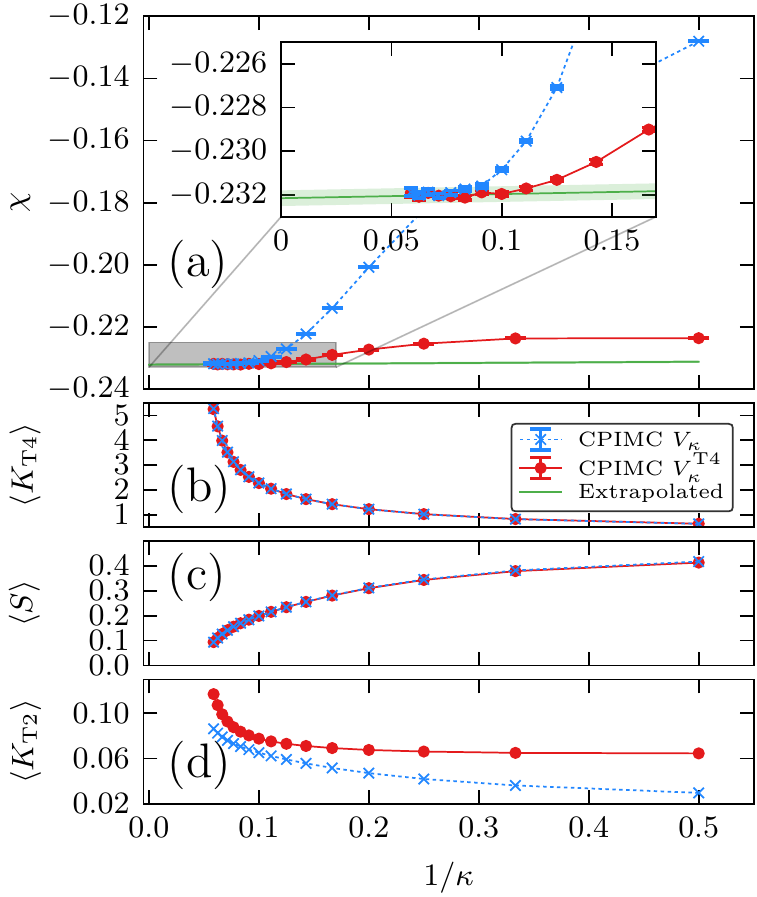}
\hspace{0.5cm}
\includegraphics[width=0.4\textwidth]{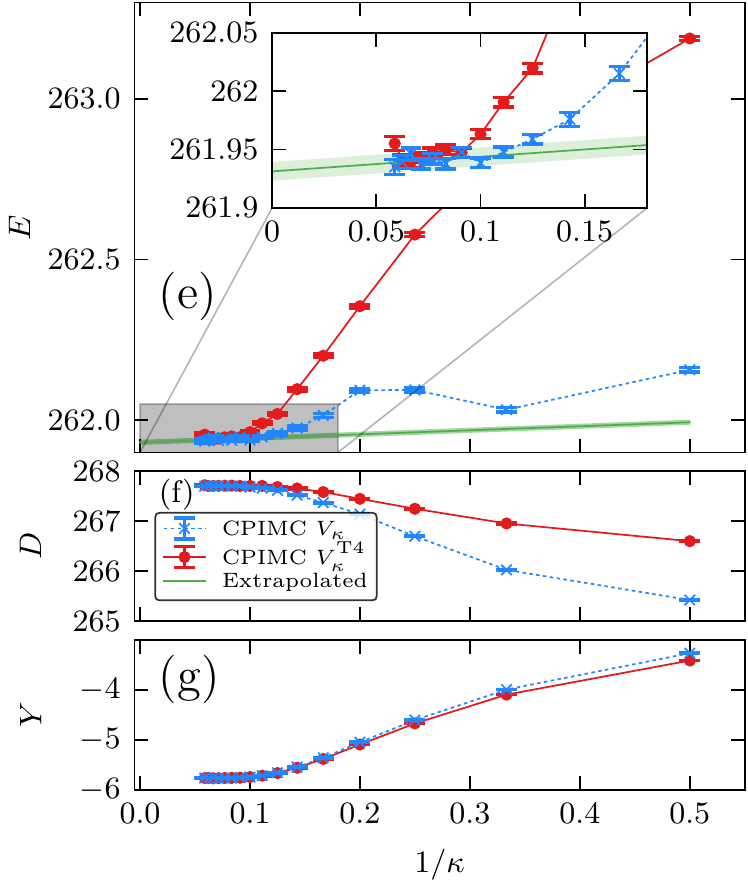}
\caption{\label{fig:fermiPot} Kink potential extrapolation of (a) the SDRF  and (e) the total energy to the exact limit, $\kappa\to\infty$. Shown are the results from CPIMC simulations of the inhomogeneous electron gas containing $N=38$ electrons at $\theta=0.5$ and $r_s=0.5$. The amplitude of the external field has been set to $A=0.2$ with a wave-vector  $\mathbf{q}=\frac{2\pi}{L}(1,1,1)^{\text{T}}$. Each data point has been obtained from a complete simulation with a fixed value for the parameter $\kappa$ in the kink potential Eq.~(\ref{eq:fermiPot}). Red points: the kink potential is applied solely to the type 4 kinks (no restriction on the number of type 2 kinks) in the sampled paths. Blue crosses: the kink potential has been applied to the total number of type 2 and 4 kinks. Green line: linear fit to the last red data points. In addition, for both potentials the dependence on the parameter $\kappa$ is plotted for the average number of type 4 kinks (b), average sign (c), type 2 kinks (d), diagonal (f), and off-diagonal contribution to the energy (g).} 
\end{figure*}

A similar observation has already been reported for the total energy of the homogeneous (unperturbed) electron gas in Ref.~\onlinecite{groth}. There, a systematic extrapolation over the maximum number of kinks (to the exact result) was conveniently realized by the use of an auxiliary kink potential:
\begin{eqnarray}\label{eq:fermiPot}
V_{\kappa}(K)=\frac{1}{e^{-(\kappa-K+0.5)}+1}\;,
\end{eqnarray}
that depends on the number of kinks $K$ of a configuration and the maximum number $\kappa$.
The procedure works as follows: the weight function $W(C)$, Eq.~(\ref{eq:weight}), is replaced  by 
\begin{equation}
W_\kappa(C) = W(C)\cdot V_{\kappa}(K),     
\end{equation}
and one performs simulations for fixed values of $\kappa$. Since it is $\lim_{\kappa\to\infty} V_{\kappa}(K)=1$, the exact partition function (and hence the exact result) is recovered by an extrapolation to $\kappa\to\infty$. This is demonstrated in Fig.~\ref{fig:fermiPot}, where we have increased the system size to $N=38$ electrons (again at $\theta=0.5$ and $r_s=0.5$). First, we focus on the blue data points, which have been obtained from a complete CPIMC simulation with a fixed value of the parameter $\kappa$ in the artificially modified weight function $W_\kappa(C)$. Here, the kink potential acts as a smooth but exponentially increasing penalty for all paths that contain a total number of kinks larger than $\kappa$. As expected, the results for the SDRF (Fig.~\ref{fig:fermiPot}. a) converge for sufficiently large $\kappa$, in this case at about $\kappa\gtrsim 10$. And since the average number of kinks [panels (b) and (d)], and consequently, the average sign [panel (c)] are clearly not converged for $\kappa \sim 10$, we can indeed conclude that all contributions from paths containing more than some critical number of kinks exactly cancel. 

We again stress that the difference between CPIMC simulations of the homogeneous and perturbed electron gas lies in the existence of type 2 kinks in the latter. In particular, the SDRF is solely computed from the type 2 kinks [cf.~its estimator, Eq.~(\ref{eq:estimator})].  In the LR regime the average number of type 2 kinks, $\braket{K_\text{T2}}$, is significantly smaller than $\braket{K_\text{T4}}$ meaning that its practical influence on the sign is negligible. Therefore, it is reasonable to apply the kink potential only to the type 4 kinks and impose no restriction on the number of type 2 kinks. Recalling that the type 4 kinks are solely due to the Coulomb correlations, this procedure is equivalent to extrapolating the true static response with respect to the correlations in the system---this procedure converges to the exact result with increasing $\kappa$. The result is shown by the red dots in Fig.~\ref{fig:fermiPot}. Evidently, the convergence with $\kappa$ is greatly accelerated. Even at $\kappa=2$, the result for the response function has only a small bias of a few percent. In contrast, when also restricting the type 2 kinks (blue crosses), the result is off by roughly a factor $2$. 

We now analyze the total energy of the inhomogeneous electron gas. Here the convergence behaviour with respect to the kink parameter $\kappa$ is different [see Fig.~\ref{fig:fermiPot} (e)]. Here, imposing no restrictions on the type 2 kinks (red points) seemingly slows down the convergence with $\kappa$. This is due to a coincidental error cancellation of the diagonal [panel (f)] and off-diagonal contributions [panel (g)] to the total energy, $E=D+Y$. Both contributions, at fixed $\kappa$, are closer to the exact result when leaving the number of type 2 kinks unrestricted. Moreover, even in the case where one is particularly interested in the total energy the potential $V_\kappa^{\text{T4}}$ (red dots) should still be used since only this potential ensures a monotonic convergence of the energy with $\kappa$. Naturally, a monotonic convergence is preferred when performing a reliable extrapolation to $\kappa\to\infty$.

From the investigations in this section we conclude that the general concept of an auxiliary kink potential to enhance the performance of CPIMC simulations that has been previously introduced for the unperturbed UEG~\cite{tim3,groth}, can be used in a similar way for the inhomogeneous electron gas. At fixed temperature and density, this allows us to obtain the SDRF for twice as large systems. This is an impressive efficiency gain when considering that the FSP increases exponentially with the system size, cf.~Eq.~(\ref{eq:fsp}). For the presented example with $\theta=0.5$ and $r_s=0.5$, CPIMC simulations without the kink potential are feasible for up to $N\sim 20$ electrons whereas, with the kink potential, simulations of $N=38$ particles pose no problem. On the other hand, for fixed temperature and electron number, use of the kink potential roughly doubles the accessible $r_s$-range, which corresponds to a factor 8 in the density. Most importantly, it turns out that, in the LR regime, the number of type 2 kinks is small compared to the number of type 4 kinks so that their practical influence on the average sign is negligible. For this reason, the accessible parameter range regarding particle number, temperature and density for which the SDRF can be computed by means of CPIMC simulations of the inhomogeneous electron gas is almost identical to the range of applicability of CPIMC to the unperturbed spatially homogeneous electron gas.

\subsection{Finite Size Correction of the Static Density Response Function\label{sec:finite_size}}

\subsubsection{Theory}
In this section, the issue of finite size errors in the computation of the SDRF $\chi$ and ways to correct them are discussed in detail. These errors are a direct consequence of the fact that Monte Carlo simulations can only be performed for a finite particle number $N$ in a finite simulation box with volume $V$. This often causes the resulting functional form of $\chi_N(\mathbf{q})$ to differ significantly from its thermodynamic limit 
\begin{eqnarray}
\chi(\mathbf{q})=\lim_{\substack{N\to \infty \\ N/V=\text{const.}}} \chi_N(\mathbf{q})\;.
\end{eqnarray}
In particular when simulating fermionic systems with Monte Carlo methods, one is usually limited to rather small systems, due to the FSP, so that finite size errors are not negligible. In addition, $\mathbf{q}-$dependent quantities can only be computed for $\mathbf{q}-$vectors that satisfy the natural condition of momentum quantization in the simulation box, $\mathbf{q}=\frac{2\pi}{L}\mathbf{m}$ with $\mathbf{m}\in\mathbb{Z}^3$. Thus, standard techniques to reduce finite size errors, e.g. those for the total energy~\cite{dornheim_prl}, which are all based on an extrapolation of the finite$-N$ results to $N\to\infty$ (at constant density) cannot be used for the correction of $\chi$.

In the ground state, the most sophisticated approach to tackle finite size errors is based on the assumption that the so-called static local field correction (LFC) $G(\mathbf{q})$ is only weakly dependent on the system size~\cite{moroni2}. The LFC is commonly defined by the equation~\cite{kugler1,kugler2}
\begin{eqnarray}\label{eq:lfc_def}
\chi(\mathbf{q})= \frac{\chi^0(\mathbf{q})}{1-v_q[1-G(\mathbf{q})]\chi^0(\mathbf{q})}\;,
\end{eqnarray}
where $\chi^0$ denotes the ideal response function and $v_q=4\pi/q^2$. The random phase approximation (RPA)~\cite{rpa_original} $\chi^\text{RPA}$ is obtained by setting $G=0$ in Eq.~(\ref{eq:lfc_def}). Hence, the LFC contains all information beyond the RPA and should thus be dominated by short-range correlations, which are expected to be captured sufficiently well in a finite simulation cell. Naturally, instead of computing the LFC from the ideal response function in the TDL, $\chi^0(\mathbf{q})$~\cite{quantum_theory}, i.e. via
\begin{eqnarray}\label{eq:G_inconsistent}
G_N(\mathbf{q}) = \frac{1}{v_q}\left(\frac{1}{\chi_N(\mathbf{q})}-\frac{1}{\chi^0(\mathbf{q})}\right) + 1\;,
\end{eqnarray}
it is important to obtain it consistently from the corresponding finite-$N$ ideal response function $\chi_N^0(\mathbf{q})$,
\begin{eqnarray}\label{eq:G_consistent}
G^\text{FSC}_N(\mathbf{q}) = \frac{1}{v_q}\left(\frac{1}{\chi_N(\mathbf{q})}-\frac{1}{\chi_N^0(\mathbf{q})}\right) + 1\;.
\end{eqnarray}
Assuming that the finite size errors in this consistent LFC are negligible, i.e., $G^\text{FSC}_N(\mathbf{q})\approx G(\mathbf{q})$, the finite size corrected response function is given by
\begin{eqnarray}\label{eq:fsc_chi}
\chi^\text{FSC}(\mathbf{q})= \frac{\chi^0(\mathbf{q})}{1+\left[\frac{1}{\chi_N(\mathbf{q})}-\frac{1}{\chi_N^0(\mathbf{q})}\right]\chi^0(\mathbf{q})}\;. 
\end{eqnarray}
Therefore, in addition to the response function of the interacting finite-$N$ system, $\chi_N(\mathbf{q})$, we also need precise data for the corresponding ideal response function $\chi^0_N(\mathbf{q})$. In principle, these can be obtained from a complete CPIMC simulation of the ideal perturbed electron gas for each $\mathbf{q}-$vector and particle number $N$, as was demonstrated in Sec.~\ref{sec:ideal}. A more convenient way to achieve this is given by making use of the spectral representation of the ideal response function, which, in case of the UEG, takes the form~\cite{quantum_theory}
\begin{eqnarray}\label{eq:ideal_chi_formula}
\chi^0_N(\mathbf{q}) = \frac{1}{V}\sum_{\mathbf{p},\sigma}\frac{n_\sigma(\mathbf{p}+\mathbf{q}) - n_\sigma(\mathbf{p})}{\epsilon_{\mathbf{p}+\mathbf{q}}-\epsilon_{\mathbf{p}}}\;,
\end{eqnarray}
\begin{figure}
\includegraphics[width=0.4\textwidth]{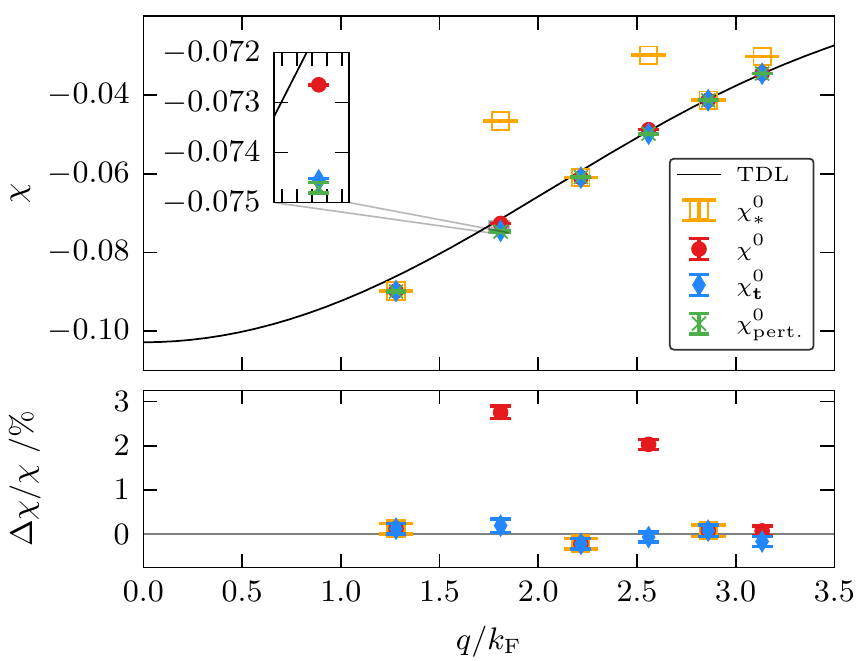}
\caption{\label{fig:ideal_correction} Comparison of different ways to compute the ideal response function for the UEG with $N=4$ electrons at $\theta=1$ and $r_s=1$. The orange squares correspond to the evaluation of Eq.~(\ref{eq:ideal_chi_formula_star}). The blue diamonds show the result from Eq.~(\ref{eq:ideal_chi_formula_twist}) for a small twist-angle $\mathbf{t}=0.01\cdot(1/e,1/\pi,1/\sqrt(2))^T$. The red dots correspond to Eq.~(\ref{eq:ideal_chi_formula_lhospital}). For comparison, the result obtained from CPIMC simulations of the perturbed ideal electron gas, as discussed in Sec.~\ref{sec:ideal}, is depicted by the green crosses. The bottom panel shows the relative deviation to this exact data. The black solid line corresponds to the ideal response function in the TDL.}
\end{figure}
\begin{figure*}
\includegraphics[width=0.3\textwidth]{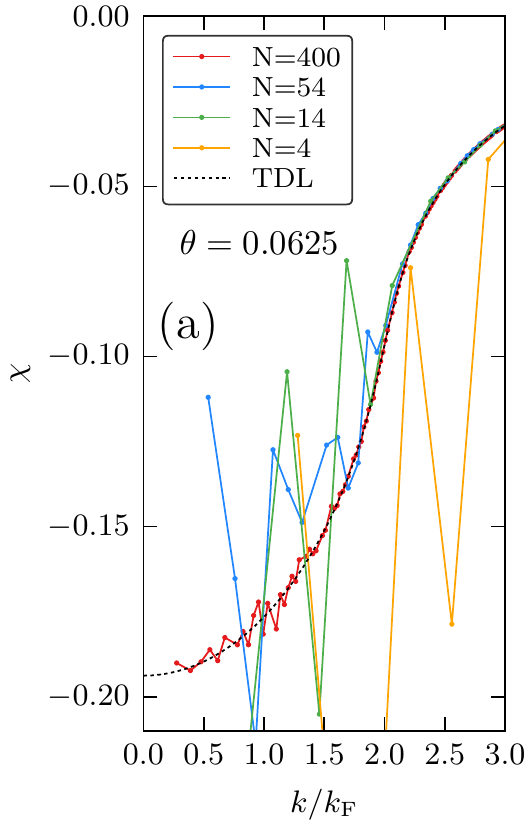}
\includegraphics[width=0.3\textwidth]{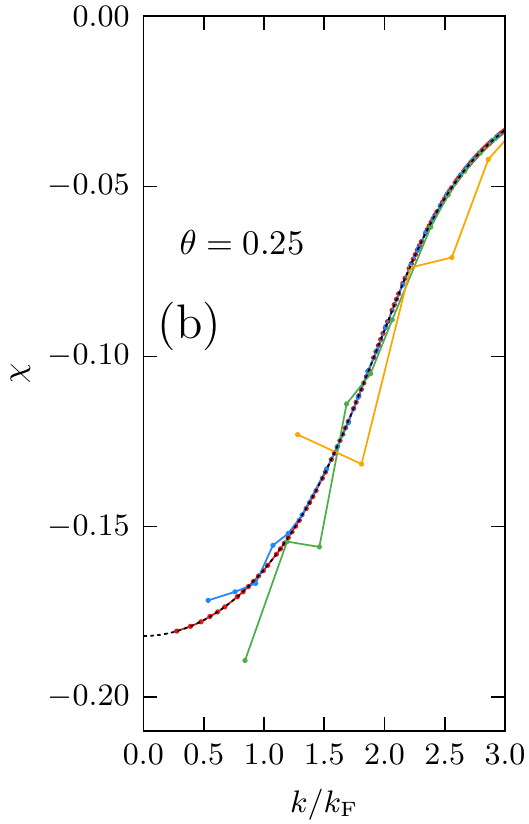}
\includegraphics[width=0.3\textwidth]{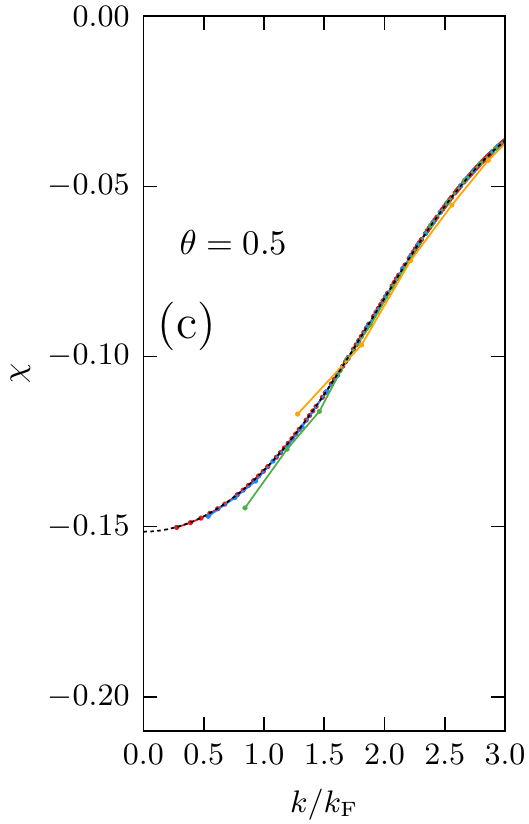}
\caption{\label{fig:ideal_N_dependence} Dependence of the ideal response function on the particle number at $r_s=1$ and three different temperatures: $\theta=0.0625$ (a), $\theta=0.25$ (b), and $\theta=0.5$ (c). The results for finite particle numbers have been computed via Eq.~(\ref{eq:ideal_chi_formula_twist}) from the CPIMC result of the corresponding finite$-N$ momentum distribution. For comparison black dotted curves show the TDL result of the ideal response function.}
\end{figure*}
where $\epsilon_{\mathbf{p}}=p^2/2$, and  $n_\sigma(\mathbf{p})=\braket{\op{n}_{\mathbf{p},\sigma}}$ is the momentum distribution of the unperturbed ideal UEG, which converges to the Fermi distribution, in the TDL, and constitutes a natural observable that is straightforwardly computed observable within the CPIMC formalism. Thus, Eq.~(\ref{eq:ideal_chi_formula}) in principle enables us to gain access to all $\mathbf{q}-$vectors of the ideal response function from a single CPIMC simulation of the unperturbed UEG. 

However, the concrete evaluation of Eq.~(\ref{eq:ideal_chi_formula}) has to be done carefully because there are terms in which both the numerator and denominator vanish, i.e., where $|\mathbf{p}+\mathbf{q}|=|\mathbf{p}|$. In the ground state it is correct to simply set those terms to zero~\cite{quantum_theory} and to rewrite
\begin{eqnarray}\label{eq:ideal_chi_formula_star}
\chi^0_{*,N}(\mathbf{q}) = \frac{1}{V}\sum_{\substack{\mathbf{p},\sigma\\ |\mathbf{p}+\mathbf{q}|\neq|\mathbf{p}|}}\frac{n_\sigma(\mathbf{p}+\mathbf{q}) - n_\sigma(\mathbf{p})}{\epsilon_{\mathbf{p}+\mathbf{q}}-\epsilon_{\mathbf{p}}}\;.
\end{eqnarray}
However, at finite temperature\cite{note} this leads to completely wrong results, which is illustrated in Fig.~\ref{fig:ideal_correction}, where the ideal response function of $N=4$ electrons at $\theta=1$ and $r_s=1$ is shown. The green crosses correspond to the exact result obtained from simulations of the perturbed ideal electron gas as discussed in Sec.~\ref{sec:ideal}. The orange squares, which correspond to the evaluation of Eq.~(\ref{eq:ideal_chi_formula_star}), exhibit a large bias for every second $\mathbf{q}-$vector, while every other is in perfect agreement with the result from the unperturbed system (see deviation in the bottom panel of Fig.~\ref{fig:ideal_correction}). This is due to the fact that the condition $|\mathbf{p}+\mathbf{q}|=|\mathbf{p}|$ can only be fulfilled if $\tilde{q}^2=q^2L^2/(2\pi)^2$ is an even number (in what follows the tilde denotes dimensionless $q-$vectors with the components $\tilde{q}_i\in\mathbb{Z}$). The proof is obvious when rewriting said condition as
\begin{align}
\tilde{p}^2 &= \tilde{p}^2+ \tilde{q}^2 + 2\tilde{\mathbf{p}}\tilde{\mathbf{q}}\\
\Leftrightarrow \tilde{q}^2 &= - 2\tilde{\mathbf{p}}\tilde{\mathbf{q}}\;.
\end{align}
Since the factor 2 ensures that the RHS is always an even number the equality can only be fulfilled if $\tilde{q}^2$ is also even. Thus, there ara no critical (diverging) terms in the evaluation of Eq.~(\ref{eq:ideal_chi_formula}) for odd $\tilde{q}^2$. 

To determine the proper contribution of the critical terms for even $\tilde{q}^2$, we may write Eq.~(\ref{eq:ideal_chi_formula}) for the UEG Hamiltonian, Eq.~(\ref{eq:H}), subject to generalized periodic boundary conditions. Following Refs.~\onlinecite{lin,drummond}, this is realized by shifting the entire $\mathbf{q}-$grid of our simulation box by a so-called twist-angle $\mathbf{t}\in\mathbb{R}^3$ so that the modified momentum quantization reads $\mathbf{q}=\frac{2\pi}{L}\mathbf{m}+\mathbf{t}$, with $\mathbf{m}\in\mathbb{Z}^3$. For the ideal response function we then have
\begin{eqnarray}\label{eq:ideal_chi_formula_twist}
\chi^0_{\mathbf{t},N}(\mathbf{q}) = \frac{1}{V}\sum_{\mathbf{p},\sigma}\frac{n_\sigma(\mathbf{p}+\mathbf{t}+\mathbf{q}) - n_\sigma(\mathbf{p}+\mathbf{t})}{\epsilon_{\mathbf{p}+\mathbf{t}+\mathbf{q}}-\epsilon_{\mathbf{p}+\mathbf{t}}}\;,
\end{eqnarray}
where, in this notation, the sum still runs over all $\mathbf{p}-$vectors with $\mathbf{p}=\frac{2\pi}{L}\mathbf{m}$, where $\mathbf{m}\in \mathbb{Z}^3$. Obviously, the condition for a vanishing denominator now reads
\begin{eqnarray}
|\mathbf{p}+\mathbf{t}+\mathbf{q}|=|\mathbf{p}+\mathbf{t}|\;,
\end{eqnarray}
which cannot be fulfilled if the components of the twist-angle $t_i$ are irrational and linearly independent, e.g. for the choice $\mathbf{t}=(1/e,1/\pi,1/\sqrt{2})^T$. In addition, for a sufficiently small modulus of the twist-angle  we can expect the induced bias to be negligible. The blue diamonds in Fig.~\ref{fig:ideal_correction} clearly show that this is indeed the case since they perfectly agree with the exact result (see bottom panel).

Finally, we determine the contribution of the critical terms in Eq.~(\ref{eq:ideal_chi_formula}) by performing the limit $|\mathbf{t}\to\mathbf{0}|$ of those terms in Eq.~(\ref{eq:ideal_chi_formula_twist}) with the aid of L'Hospitals's rule yielding:
\begin{eqnarray}\label{eq:ideal_chi_formula_lhospital}
\chi^0_{N}(\mathbf{q}) =\chi^0_{*,N}(\mathbf{q})  
-\frac{\beta}{V}\sum_{\substack{\mathbf{p},\sigma\\ |\mathbf{p}+\mathbf{q}|=|\mathbf{p}|}} \left[n_\sigma(\mathbf{p})-n^2_\sigma(\mathbf{p})\right]\;.\quad
\end{eqnarray}
The corresponding result is depicted by the red dots in Fig.~\ref{fig:ideal_correction}. Evidently, compared to simply omitting the contribution of the critical terms (orange squares) the improvement is substantial. Yet, the relative deviation to the exact result is still of the order of a few percent (bottom panel). 

The residual bias is explained as follows: mathematically it is only valid to use L'Hospital's rule if the functional form of the momentum distribution does not change with the twist-angle. This condition only holds in good approximation for large particle numbers, but is increasingly violated for smaller system sizes. Since a systematic error of a few percent in the ideal response function is not sufficient for a reliable finite size correction, we conclude that Eq.~(\ref{eq:ideal_chi_formula_lhospital}) cannot be used to achieve this. Nevertheless, we can instead use Eq.~(\ref{eq:ideal_chi_formula_twist}), which has been demonstrated to be asymptotically correct for small twist-angles, to efficiently compute the finite-$N$ ideal response function of the UEG with high accuracy. For completeness we mention that the L'Hospital terms vanish in the ground state, and the functional form of the momentum distribution is independent of the twist-angle here, since it is always given by a step function at the Fermi vector $k_\text{F}$. Hence, Eq.~(\ref{eq:ideal_chi_formula_star}) is indeed correct in the ground state.

\begin{figure*}
\includegraphics[width=0.65\textwidth]{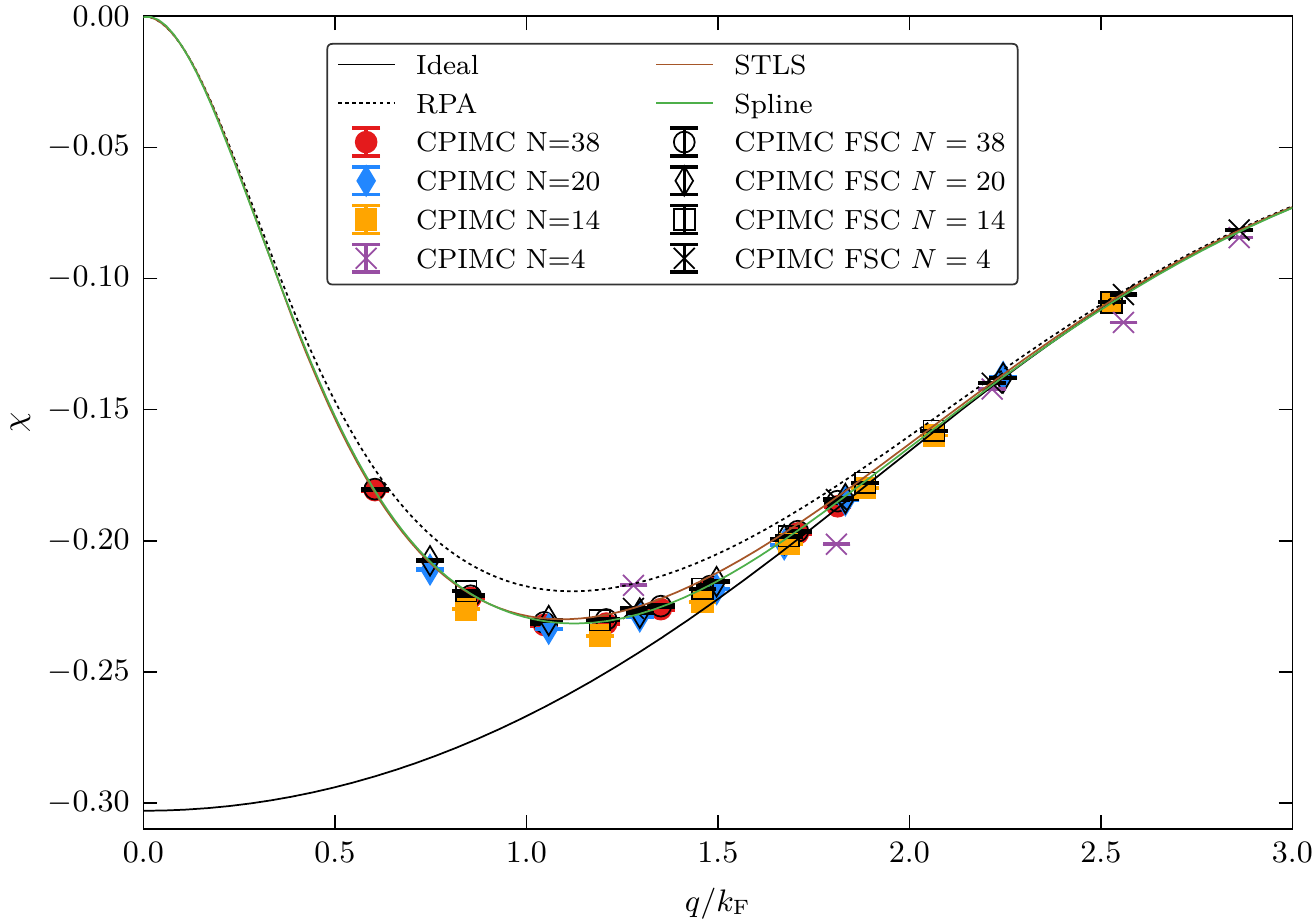}
\includegraphics[width=0.65\textwidth]{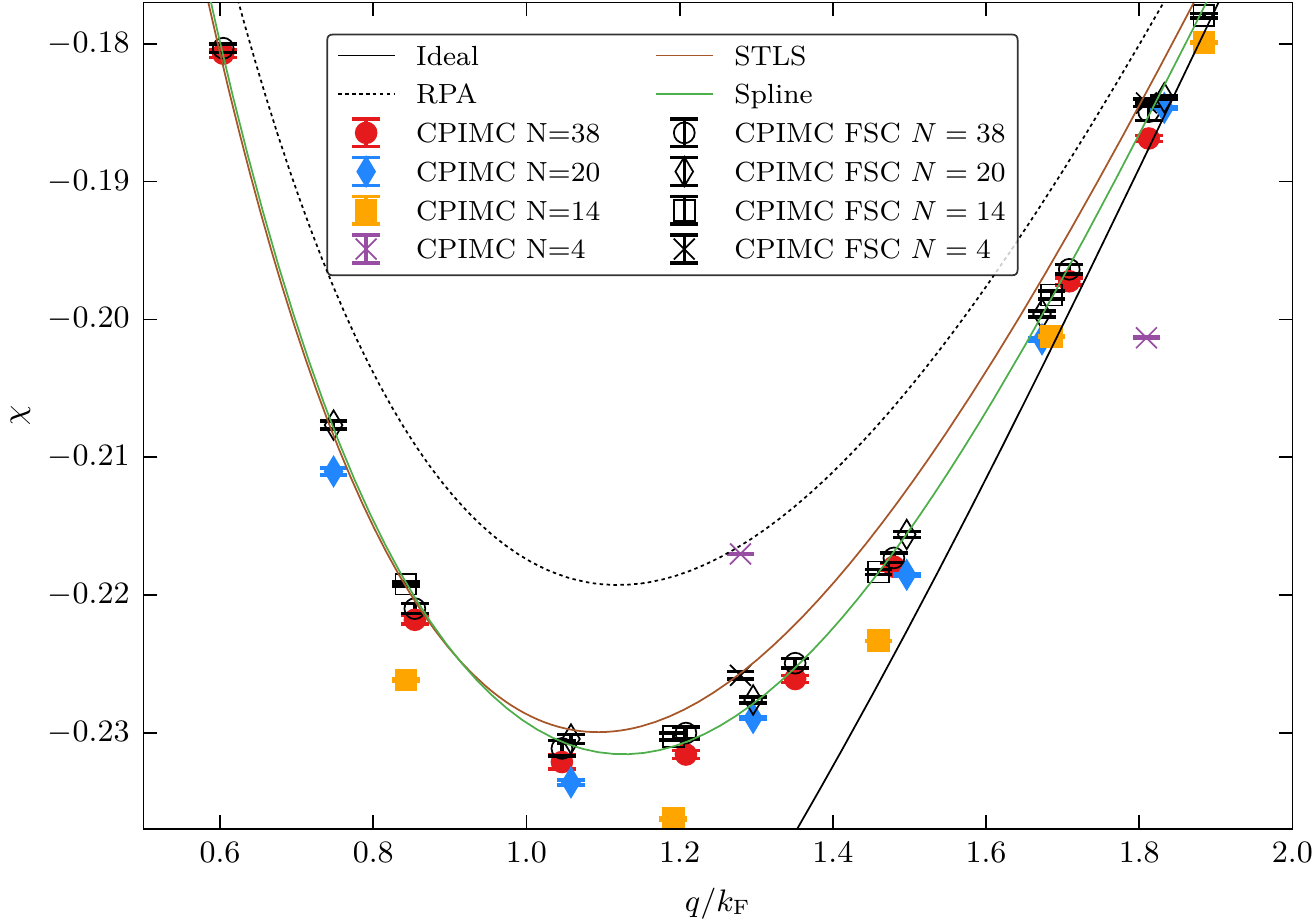}
\caption{\label{fig:total_chi} Finite size correction of the density response function of the UEG at $\theta=0.5$ and $r_s=0.5$. \textbf{Top panel:} Shown are the uncorrected CPIMC results for different electron numbers in the simulation box: 4 (purple crosses), 14 (orange squares), 20 (blue diamonds), and 38 (red dots). The black symbols correspond to the finite size corrected results computed via Eq.~(\ref{eq:fsc_chi}), and the green curve shows a smooth spline fit through these data with $N>4$. For comparison, the ideal (solid black), RPA (dotted black), and STLS (brown) results are plotted. \textbf{Bottom panel:} Zoom into the minimum regime of the response function.}
\end{figure*}
\begin{figure}
\includegraphics[width=0.49\textwidth]{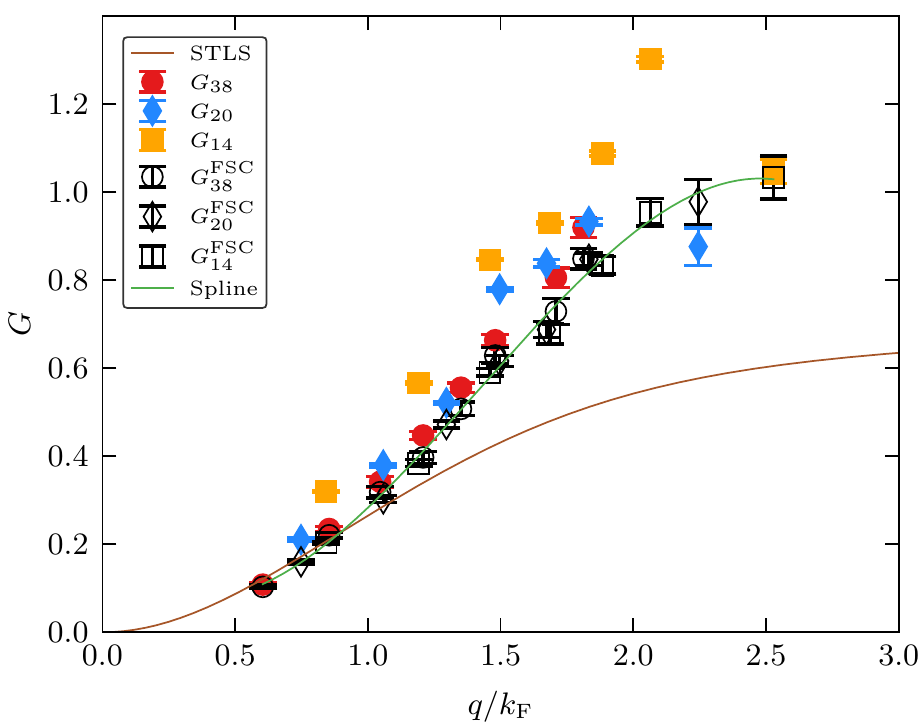}
\caption{\label{fig:LFC} Local field correction of the UEG at $\theta=0.5$ and $r_s=0.5$ for different particle numbers indicated in the legend (subscripts). Colored filled symbols: LFC computed from the ideal response function in the TDL,~Eq.~(\ref{eq:G_inconsistent}). Black symbols: LFC obtained from the finite$-N$ ideal response function according to Eq.~(\ref{eq:G_consistent}). Green curve: spline fit to the finite size corrected LFC. Brown curve: STLS local field correction.}
\end{figure}
\subsubsection{CPIMC results}
At high densities, we expect the finite size errors involved in the response function of the interacting system to be comparable to those of the ideal system. Therefore, Fig.~\ref{fig:ideal_N_dependence} shows the dependence of the ideal response function on the particle number at three different temperatures. At $\theta=0.0625$ [panel (a)], which is close to the ground state, the finite size errors are extremely large even for $N=54$ electrons (blue), and are most pronounced for small $\mathbf{q}-$vectors, which correspond to large distances in real space that are not sufficiently described in small simulation cells. It is only at a few hundred electrons (red) where the convergence of the functional form eventually becomes visible. With increasing temperature these finite size errors are significantly reduced; yet the relative bias of e.g. $N=14$ electrons at $\theta=0.5$ [green dots in panel (c)] is still substantial. This reduction of finite size errors is due to the fact that shell effects, which also cause quantities like the total energy to converge non-monotonically towards the TDL, vanish with increasing temperature.

Finally, in Fig.~\ref{fig:total_chi} the wave-vector dependence of the interacting response function of the UEG is depicted for $\theta=0.5$ and $r_s=0.5$. The colored symbols show the uncorrected CPIMC results for N=4,14,20, and 38 electrons, which have been obtained as discussed in Sec.~\ref{sec:interacting}. For $N=38$, the extrapolation technique with the kink potential has been used. First, we clearly see that the uncorrected results do not lie on a smooth curve. In particular for $N=4$ and $N=14$ electrons the finite size errors are of the order of a few percent when zooming into the minimum region of the response function (bottom panel). Before applying the presented finite size correction to $\chi$, we check if the underlying assumption regarding the weak finite size dependence of the LFC is actually valid. For this purpose, in Fig.~\ref{fig:LFC} we plot the LFC of the UEG for the same parameters. Evidently, using the ideal response function in the TDL to compute the LFC according to Eq.~(\ref{eq:G_inconsistent}) leads to substantial finite size errors in its functional form. However, when consistently using our computed CPIMC result for the finite$-N$ ideal response function (cf.~Eq.~(\ref{eq:G_consistent})), the functional form of the LFC is indeed indistinguishable for all three particle numbers so that a smooth spline can be fitted through these data (green line). For comparison, we also plot the LFC obtained from the Singwi-Tosi-Land-Sj\"olander (STLS) scheme, which is of good quality for $q/k_\text{F}\lesssim 1$ but deviates by up to a factor of two from the exact CPIMC result, for larger $q-$vectors. 

Now we use the consistent LFC to correct the SDRF according to Eq.~(\ref{eq:fsc_chi}). The result is shown by the black symbols in Fig.~\ref{fig:total_chi}. Clearly, for $N>4$ all results  lie on a smooth curve, which is demonstrated by a smooth spline fit through these data (green curve). Even though for $N=4$ (black crosses) the correction is not quite sufficient to describe the TDL behavior, the reduction of the bias is still impressive (cf.~purple crosses). In addition, we plot the response function in RPA (dotted black) and STLS (solid brown) approximation. While the RPA exhibits systematic errors of a few percent the STLS approximation is accurate up to about one percent. In particular, STLS exhibits no resolvable bias for $q/k_\text{F}\lesssim 1$, which is in agreement with its accuracy regarding the LFC (cf.~Fig~\ref{fig:LFC}) in this regime. However, even though at $q/k_\text{F}\gtrsim 2$ the systematic error of the STLS result for the LFC is nearly a factor two, the influence of the LFC on the total response function is suppressed by the factor $v_q=4\pi/q^2$ in Eq.~(\ref{eq:lfc_def}) so that for $q\to\infty$ the response function becomes equal to the ideal case. 

We conclude that the ground state finite size correction of the LFC and the SDRF can be generalized to finite temperatures, as presented in this section. The benefit of this correction is dramatic:  it allows one to obtain accurate results for  the thermodynamic limit from CPIMC simulations for systems as small as $N=14$ electrons. The price one has to pay is to compute highly accurate results of the finite$-N$ ideal response function. This can be efficiently achieved via CPIMC simulations of the unperturbed UEG by using its spectral representation. However, in contrast to the ground state, the correct evaluation of the spectral representation is only possible when switching to a system subject to generalized boundary conditions, which has been verified by cross-checks to the exact result obtained from simulations of the perturbed electron gas. Finally, we mention that the presented finite size correction is not only highly valuable for CPIMC but can be used for finite$-N$ data obtained with any other finite temperature method.

\section{Summary and Discussion\label{sec:summary}}
In summary, we have successfully generalized the CPIMC formalism from the  homogeneous electron gas to the general inhomogeneous case. We have shown that the applied external periodic potential results in the occurrence of  type 2 kinks that correspond to one-particle excitations in the simulated imaginary time paths. This leads to numerous additional diagrams, which have to be taken into account, so that the complexity of the algorithm is significantly increased. Next, we have demonstrated that the technique of an artificial kink-potential, which had been introduced in Refs.~\onlinecite{tim3,groth} to mitigate the FSP regarding the computation of the energy of the UEG, is similarly effective for the computation of the SDRF. This concept may even be improved when being applied solely to the type 4 kinks while imposing no restrictions on the type 2 kinks. Interestingly, we observed that the induced type 2 kinks only influence the fermion sign problem of CPIMC for large amplitudes of the external potential. For amplitudes that are sufficiently small for the linear response theory to be valid their influence is negligible. Therefore, the presented CPIMC algorithm can be used to compute the SDRF for the same parameters (density, temperature and electron number) that are accessible for the simulation of the UEG without the external potential.

A further achievement of this work consists in the extension of ground state finite size corrections for the SDRF to finite temperature. We have demonstrated that the SDRF obtained from quantum Monte-Carlo simulations of finite systems, i.e., a finite number of electrons in a finite simulation box, may differ substantially from the TDL result. For the investigated example of intermediate temperature ($\theta=0.5$) and rather high density ($r_s=0.5$) the finite size errors are of the order of several percent. Similarly to previous findings in the ground state, the finite size effects are almost exclusively ascribed to the ideal part of the SDRF, whereas the LFC is remarkably well converged with system size even for small $N$, i.e.~$G_N^\text{FSC}(q)\approx G(q)$. 

To compute $G^\text{FSC}_N$ from the QMC data for the SDRF we found that it is crucial to use the ideal SDRF for the same finite number of electrons (instead of using the macroscopic result), which turns out to be surprisingly difficult. While the finite$-N$ ideal SDRF is linked to the momentum distribution function via its spectral representation, at finite temperature, the corresponding expression can only be evaluated when introducing generalized boundary conditions by means of a finite but small twist-angle. Thereby, unbiased results for the finite$-N$ ideal SDRF for all wave-vectors can be obtained from a single CPIMC simulation of the unperturbed UEG. This has been confirmed by cross-checks with the exact results from simulations of the perturbed UEG. In this way, the SDRF can be computed in the TDL with an accuracy of $\sim 0.1\%$. Finally, our \textit{ab initio} results for the SDRF allow us to benchmark standard approximations. In particular the RPA SDRF~\cite{rpa_original} reveals systematic errors of a few percent, while the STLS approximation~\cite{stls_original,stls} exhibits deviations of up to one percent, even at $r_s=0.5$. 

We expect the presented results to be of high importance for future warm dense matter research, in particular in the context of advanced truly non-local exchange-correlation functionals for DFT or as valuable input for the computation of the dynamic structure factor, e.g., within the extended Born-Mermin approach\cite{fortmann2}.


\section*{Acknowledgements}
This work was supported by the Deutsche Forschungsgemeinschaft via project BO1366-10 and via SFB TR-24 project A9 as well as grant shp00015 for CPU time at the Norddeutscher Verbund f\"ur Hoch- und H\"ochstleistungsrechnen (HLRN).

\section{Appendix}
\begin{figure*}
\includegraphics[width=0.7\textwidth]{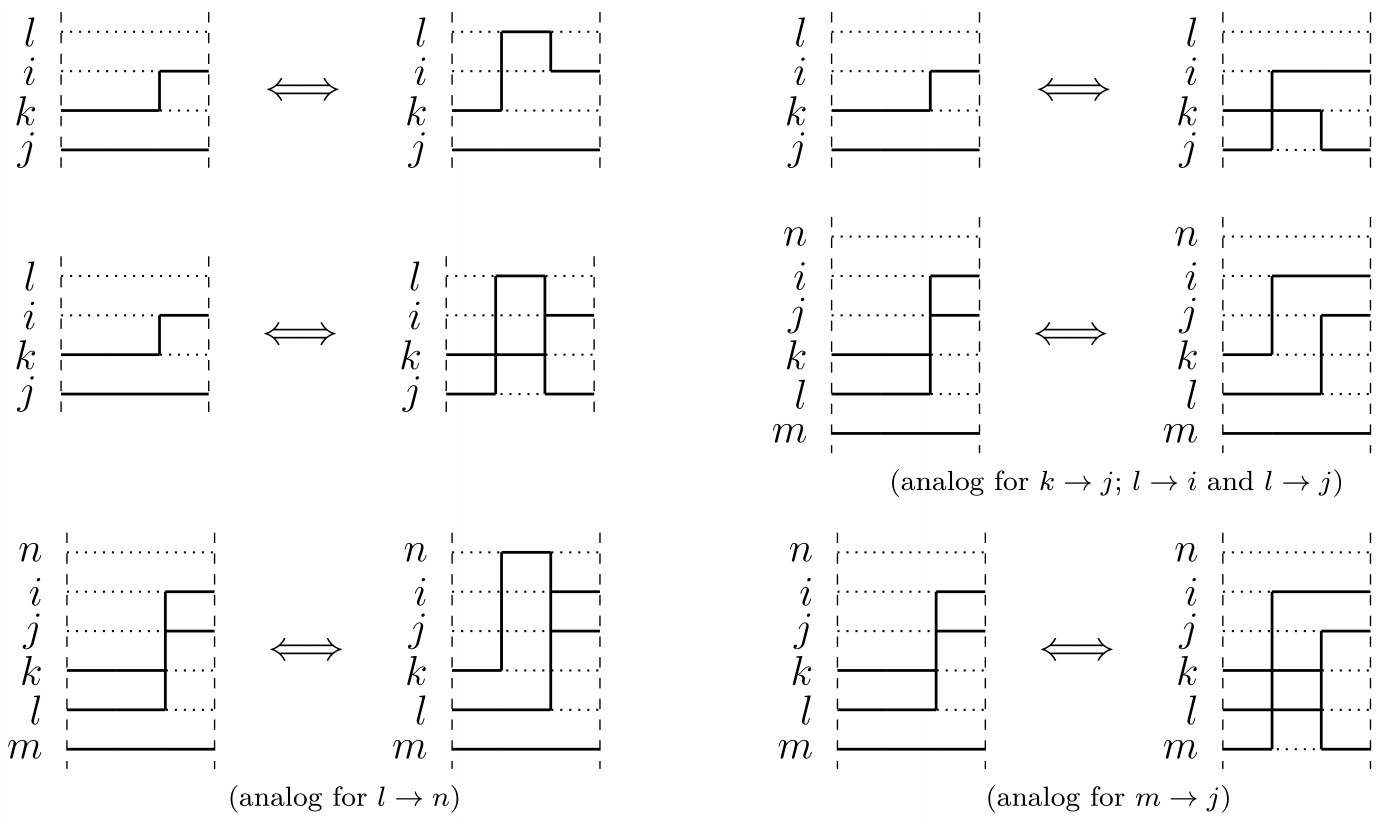}

\includegraphics[width=0.7\textwidth]{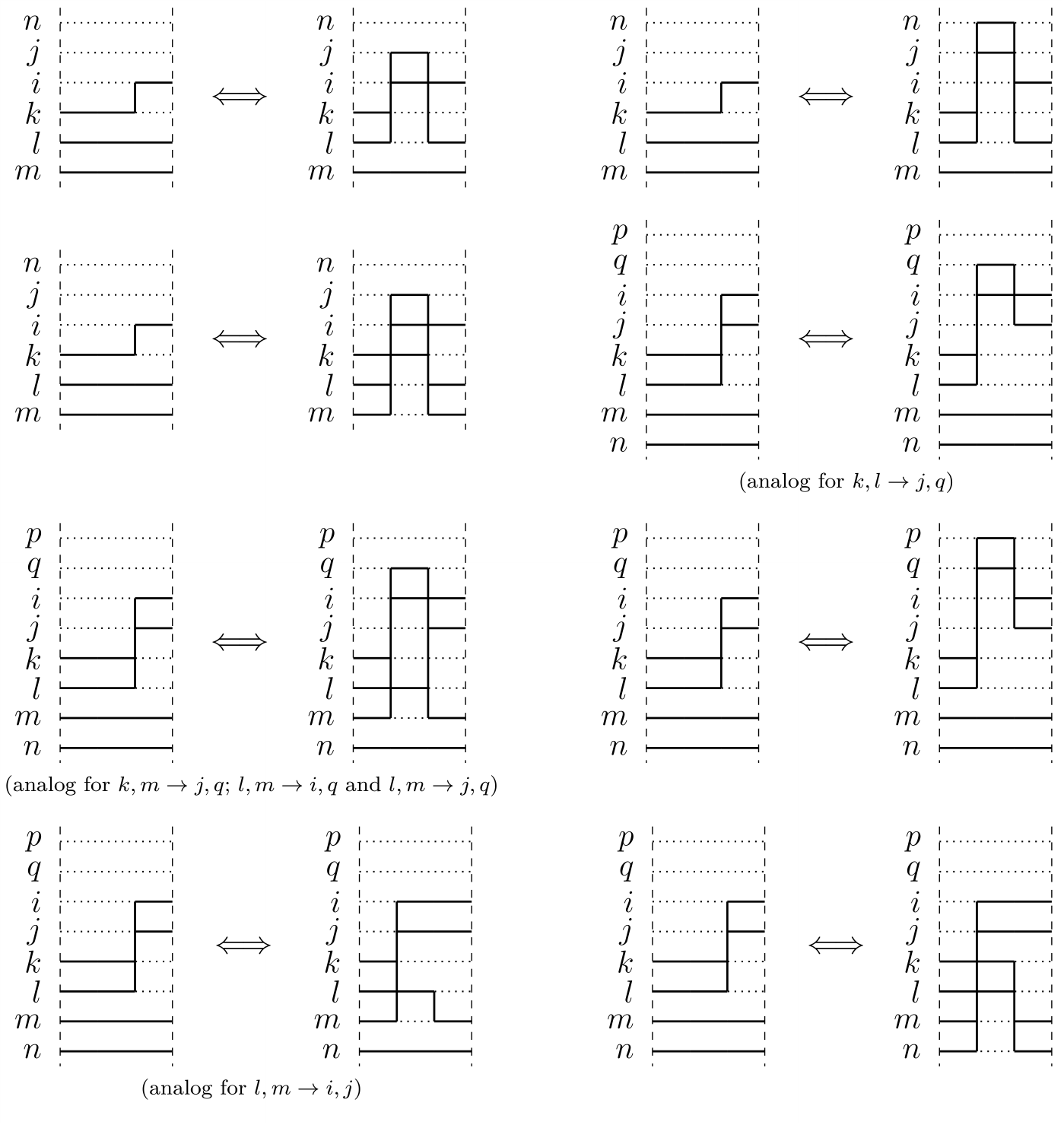}
\caption{\label{fig:diagram_add_T2T4} All 14 elementary diagrams for adding a type 2 or 4 kink via a one- or two-particle excitation, respectively, and thereby changing another kink left of the added one.}
\end{figure*}

In this appendix we present additional information on the CPIMC procedure for the harmonically modulated electron gas.
Figure~\ref{fig:diagram_add_T2T4} shows all possible 14 elementary diagrams for adding a type 2 or 4 kink via a one- or two-particle excitation, and thereby changing another kink left of the added one.

\onecolumngrid
\clearpage
\twocolumngrid


\begin{thebibliography}{26}
\bibitem{fermi_E} In Hartree atomic units, it holds $E_\text{F}=k_\text{F}^2/2$, with $k_\text{F}=(9\pi/4)^{1/3} / r_s$ being the Fermi wave vector.
\bibitem{note} Note that the commonly used expansions around $\mathbf{q}=0$ cannot be used since the denominator can vanish for arbitrarily large $\mathbf{q}$.
\end{thebibliography}
\end{document}